\documentclass[twocolumn,aps,prl,amsmath,amssymb,superscriptaddress,longbibliography]{revtex4-2}

\usepackage{newtxtext}
\usepackage{newtxmath}
\usepackage{microtype}

\usepackage{graphicx}
\usepackage{dcolumn}
\usepackage{bm}
\usepackage{multirow}
\usepackage{xcolor}
\usepackage{tikz}
\usetikzlibrary{arrows.meta}
\definecolor{PRLBlue}{RGB}{46,48,146}
\definecolor{TRSHeader}{HTML}{273B52}
\definecolor{TRSRowLight}{HTML}{F3F6FA}
\definecolor{TRSRowShade}{HTML}{E5ECF4}
\definecolor{TRSAccent}{HTML}{23588C}
\newcounter{mainbookmark}
\newcommand{\runinsection}[2][0.6\baselineskip]{%
    \par\addvspace{#1}%
    \stepcounter{mainbookmark}%
    \pdfbookmark[1]{#2}{main.section.\arabic{mainbookmark}}%
    \noindent\mbox{\color{PRLBlue}\itshape #2}\hspace{0.15em}---\hspace{0.3em}\ignorespaces
}
\usepackage{braket}
\usepackage{CJK}
\usepackage{indentfirst}
\usepackage{cases}

\usepackage{mathtools}

\def\Eq#1{Eq.~(\ref{#1})}

\newcommand{\bea}{\begin{eqnarray}}
\newcommand{\eea}{\end{eqnarray}}

\newcommand{\Tr}{\operatorname{Tr}}
\newcommand{\diag}{\operatorname{diag}}
\newcommand{\dd}{\mathrm{d}}
\newcommand{\ee}{\mathrm{e}}
\newcommand{\ii}{\mathrm{i}}
\newcommand{\trans}{\mathsf{T}}
\newcommand{\hc}{\dagger}
\newcommand{\mc}[1]{\mathcal{#1}}
\newcommand{\bb}[1]{\bm{#1}}

\newtheorem{theorem}{Theorem}

\usepackage[
    colorlinks,
    citecolor=PRLBlue,
    linkcolor=PRLBlue,
    urlcolor=PRLBlue,
    bookmarksopen=true,
    bookmarksopenlevel=1,
    bookmarksdepth=2,
]{hyperref}

\newcommand{\ror}[2]{\href{https://ror.org/#1}{#2}}

\usepackage{orcidlink}
\usepackage{lipsum}

\makeatletter
\let\oldaddcontentsline\addcontentsline
\renewcommand{\addcontentsline}[3]{}
\makeatother

\begin{document}

\title{Sign-problem-free Bosonic Auxiliary Field Quantum Monte Carlo}

\author{Yin-Kai Yu\,\orcidlink{0009-0006-2589-2772}}
\affiliation{Beijing National Laboratory for Condensed Matter Physics\\ \& \ror{05cvf7v30}{Institute of Physics, Chinese Academy of Sciences}, Beijing 100190, China}
\affiliation{\ror{05qbk4x57}{University of Chinese Academy of Sciences}, Beijing 100049, China}

\author{Zhong-Chao Wei\,\orcidlink{0000-0002-1708-4306}}
\email{zcwei@thp.uni-koeln.de}
\affiliation{Qingdao Binhai University, Qingdao 266555, China}

\author{Zi-Xiang Li\,\orcidlink{0000-0002-3941-365X}}
\email{zixiangli@iphy.ac.cn}
\affiliation{Beijing National Laboratory for Condensed Matter Physics\\ \& \ror{05cvf7v30}{Institute of Physics, Chinese Academy of Sciences}, Beijing 100190, China}
\affiliation{\ror{05qbk4x57}{University of Chinese Academy of Sciences}, Beijing 100049, China}

\date{\today}

\begin{abstract}
    The sign problem poses a fundamental barrier to numerically exact quantum Monte Carlo simulations, severely hindering our understanding of quantum many-body physics across disciplines ranging from condensed matter to particle and nuclear physics. Geometrical frustration makes this barrier especially acute for interacting bosons. Here, we develop a bosonic auxiliary-field quantum Monte Carlo (BAFQMC) method and rigorously prove that time-reversal symmetry and reflection positivity each provide a general sufficient condition for sign-problem-free simulations. Crucially, under either criterion, BAFQMC evades the sign problem even on frustrated lattices where conventional world-line and stochastic-series-expansion methods fail. Benchmarking our approach on a two-flavor Bose-Hubbard model on the triangular lattice yields excellent quantitative agreement with finite-temperature exact diagonalization. By establishing unifying principles to solve the sign problem in BAFQMC, our work substantially expands the reach of unbiased, numerically exact simulations for frustrated bosonic matter.
\end{abstract}

\maketitle
\pdfbookmark[0]{Main Text}{main}

\phantomsection\label{sec:introduction}
\runinsection[0pt]{Introduction} Understanding interacting quantum many-body systems remains a central challenge in modern physics~\cite{sachdev2011qpt,fradkin2013ftcmp}. Quantum Monte Carlo (QMC) methods are among the major numerical approaches for quantitative studies of these systems, particularly in more than one spatial dimension~\cite{blankenbecler1981prd,scalapino1981prl,hirsch1981prl,hirsch1985prb,sugiyamaKoonin1986,prokofev1998pla,sandvik1999prb}. They are invaluable for strongly correlated matter, delivering unbiased and numerically exact results where perturbative frameworks often fail. However, the central obstacle of QMC simulations is the notorious sign problem: when quantum-mechanical configuration weights are not non-negative, the required computational effort scales exponentially with system size and inverse temperature~\cite{loh1990prb}. Therefore, the sign problem represents a fundamental barrier to exploring quantum many-body physics in various areas, ranging from condensed matter to particle and nuclear physics~\cite{li2019arcmp}. Although no universal solution is expected~\cite{troyer2005prl}, the sign problem can be avoided in important classes of physical systems. Indeed, much of QMC's continued impact~\cite{burovski2006prl,sandvik2007prl,capogrossoSansone2008pra,berg2012science,nasu2015prb,otsuka2016prx,Sandvik2016science,Huang2017science,Zhang2018prl,teixeira2023prb,Ding2026arxiv} rests on the ongoing discovery of sign-free formulations for pivotal models~\cite{lang1993prc,wu2005prb,li2015prb,li2016prl,chandrasekharanWiese1999,wang2015prl,wei2016prl,niuLu2025,Han2024pfaffianquantummontecarlo}, alongside a deepening theoretical understanding of the sign problem that has inspired novel strategies to avoid or mitigate it~\cite{li2019arcmp,hangleiter2020,levyClark2021,guptaAlbashHen2020,wanYao2022,mondainiTaratScalettar2022,zhangPanXuMeng2022,Mondaini2023prb,Li2022arXivAsymtotic,Berg2023PRL,yu2026,Qin2026PRB,Zhang2025PRB,LiXZ2026Signproblem,Shackleton2026PRL,Mondaini2026arxiv}.

For bosonic systems, the sign problem is commonly attributed to geometrical frustration~\cite{hatanoSuzuki1992,henelius2000prb}. Off-diagonal hopping or exchange processes carry signs that cannot generally be made mutually compatible around closed loops of frustrated lattices such as the triangular and kagome lattices~\cite{hatanoSuzuki1992,henelius2000prb,hen2021}. Apart from a limited number of models where specially designed formulations or regroupings eliminate or substantially reduce the negative signs~\cite{nakamura1998prb,nyfeler2008,alet2016prl,wessel2017,stapmanns2018,hangleiter2020,weber2022,weberCaciWessel2022,murotaTodo2025,Shackleton2026PRL,Herz2026arxiv}, most frustrated bosonic systems still lack sign-problem-free QMC formulations. For broad classes of bosonic topological phases, the sign problem is even believed to be intrinsic~\cite{hastings2016,ringelKovrizhin2017,golanSmithRingel2020,smithGolanRingel2020,seo2025}. Geometrical frustration has therefore become almost synonymous with the sign problem in bosonic QMC. This computational barrier has placed some of the most sought-after phenomena in quantum matter beyond the reach of unbiased large-scale simulations, including bosonic fractional quantum Hall states~\cite{sorensen2005prl,wang2011prl,leonard2023}, frustrated supersolids~\cite{wang2009prl,gallegos2025}, chiral Mott phases and the associated quantum criticality~\cite{dhar2013,zaletel2014prb,baldelli2024}, frustrated superfluids~\cite{you2012prl,jiangFuXu2012,burba2026}, and quantum spin liquids~\cite{yan2011science,zhu2015prb,zhou2017rmp,balents2010,gallegos2025}.

However, this familiar identification of frustration with the sign problem in bosonic systems is rooted in a specific paradigm: QMC frameworks that assemble statistical weights from local matrix elements, such as world-line QMC (WLQMC) and stochastic-series-expansion (SSE) methods~\cite{suzuki1976,sandvik1999prb,prokofev1998pla,evertz2003advphys,syljuasenSandvik2002,satoAssaadGrover2018,ringelKovrizhin2017,Rayudu2025arxiv}. By contrast, auxiliary-field QMC (AFQMC)---the workhorse algorithm for interacting fermions---circumvents this local construction entirely. Through a Hubbard-Stratonovich (HS) transformation, interactions are decoupled and the original quantum degrees of freedom are traced out exactly at fixed auxiliary fields, shifting the Monte Carlo sampling to the resulting global weight~\cite{stratonovich1957,hubbard1959,scalapino1981prl,blankenbecler1981prd,hirsch1981prl,assaadEvertz2008,Assaad2025scipostcodebases,CohenStead2026scipostcodebases}. Crucially, because the local quantum degrees of freedom are traced out, geometrical frustration of individual matrix elements no longer dictates the sign~\cite{satoAssaadGrover2018}. For fermions, while AFQMC can still suffer from negative determinant weights~\cite{loh1990prb}, the global symmetries and mathematical structures of the HS-decoupled Hamiltonian can rigorously protect weight positivity across broad classes of models. Principles such as time-reversal symmetry (TRS)~\cite{lang1993prc,wu2005prb,li2015prb,li2016prl}, reflection positivity (RP)~\cite{wei2016prl}, and related algebraic structures~\cite{wang2015prl,wei2024prb,li2019arcmp} have provided general routes to sign-problem-free simulations, enabling a wealth of large-scale fermionic studies~\cite{berg2012science,Wang2014prl,schattner2016prx,li2017prb,schattner2016prl,liJiangJianYao2017natcomm,jiang2017prb,gazit2017natphys,assaadGrover2016prx,gerlach2017prb,liJiangYao2017prl,lederer2017pnas,YU2024PRL,liVaeziMendlYao2018sciadv,liu2019natcomm,He2019PRL,Xu2023prr,Grover2021PRL,hofmann2022prx,liu2022prl,hofmann2023prl,lunts2023natcomm,Yu2024PRL2,cai2025pnas,patel2025prx,He2025PRL,wang2026arxivsuperradiantstronglycorrelatedquantum,Xu2026arXiv,shi2026arxivhightemperaturecharge4esuperconductivitysu4,He2026arXiv}. Although AFQMC and related methods have also been applied to interacting bosons~\cite{sugiyamaKoonin1986,ulmkeScalettar2000,purwanto2005pra,purwanto2004pre,Rubenstein2012PRA}, no unifying criteria currently exist to guarantee sign-free weights. Consequently, whenever negative or complex weights appear, exact bosonic calculations remain confined to small systems, forcing large-scale simulations to rely on systematically biased approximations~\cite{purwanto2004pre,Rubenstein2012PRA,malpettiRoscilde2017,Cai2025CPB}. It is therefore both crucial and urgent to establish general principles, rooted in the Hamiltonian's global symmetries and mathematical structures, that can systematically solve the sign problem in bosonic systems.

\begin{figure}[bt]
    \centering
    \includegraphics[width=\linewidth]{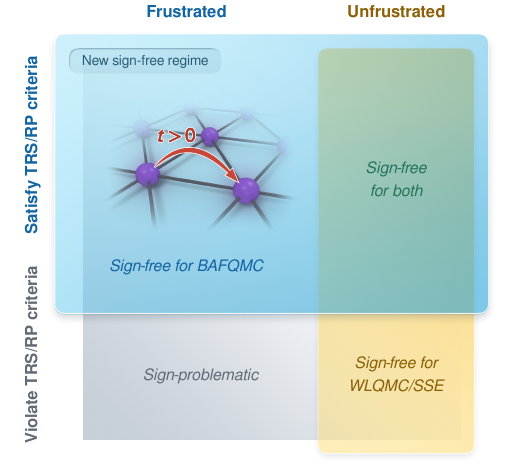}
    \caption{\textbf{Schematic illustration of sign-problem-free regimes in bosonic QMC.} Frustrated and unfrustrated lattices occupy the left and right regions. The upper region satisfies the time-reversal symmetry (TRS) or reflection positivity (RP) criterion for the HS-decoupled Hamiltonian. BAFQMC is sign-free throughout this region, while world-line QMC (WLQMC) and stochastic-series-expansion QMC (SSE) are sign-free in the right region. The upper-left quadrant marks the new sign-free regime unlocked by this work, illustrated by frustrated nearest-neighbor hopping ($t>0$) on a triangular lattice. By tracing out bosons at fixed auxiliary fields, BAFQMC shifts the basis for positivity from local matrix elements to global TRS/RP protection of the resulting weights, opening frustrated bosonic systems to sign-free QMC.}
    \label{fig:schematic}
\end{figure}

In this Letter, we provide unifying criteria for sign-problem-free AFQMC simulations in bosonic systems. We introduce a finite-temperature bosonic auxiliary-field quantum Monte Carlo (BAFQMC) method and rigorously prove that TRS and RP each furnish a general sufficient condition for strictly positive configuration weights. As illustrated in Fig.~\ref{fig:schematic}, whenever the HS-decoupled Hamiltonian satisfies either of these criteria, our BAFQMC framework completely evades the sign problem, even in strongly frustrated lattice geometries. This development brings unbiased, numerically exact simulations of a broad class of frustrated models within reach, bypassing the fundamental limitations of WLQMC and SSE and opening a new avenue to explore long-sought phases of frustrated bosonic matter. To benchmark the method and demonstrate its reliability, we solve a frustrated two-flavor Bose-Hubbard model on the triangular lattice, finding excellent quantitative agreement with finite-temperature exact diagonalization.

\phantomsection\label{sec:afqmc}
\runinsection{Bosonic auxiliary-field quantum Monte Carlo}
In AFQMC, interactions are decoupled by a Hubbard-Stratonovich (HS) transformation~\cite{stratonovich1957,hubbard1959}. The Trotter and HS decompositions introduce spacetime auxiliary fields $\phi$, and for each fixed field configuration the bosons evolve under one-body Hamiltonians. BAFQMC samples these auxiliary fields and evaluates the bosonic trace exactly at fixed $\phi$. The partition function takes the form
\begin{equation}
    Z = \int \mc D\phi\; P_{\rm HS}(\phi)\, W(\phi),
    \quad
    W(\phi)=\Tr\left[\prod_{\ell=1}^{L}
    \ee^{\hat h_\ell(\phi_\ell)}\right].
    \label{eq:Z_afqmc}
\end{equation}
Here $L$ is the number of factors in the ordered product, $\ell$ labels the spacetime auxiliary-field variables generated by the Trotter and HS decomposition, and $P_{\rm HS}(\phi)=\prod_{\ell}\frac{1}{\sqrt{2\pi}}\ee^{-\frac12\phi_{\ell}^{2}}$ is the product of the Gaussian factors introduced by the HS transformation~\cite{SM}. The sign problem in AFQMC is that $W(\phi)$ is not necessarily a positive real number over auxiliary-field configurations.

The simplest case is a one-body evolution that conserves particle number. Each slice is then a bosonic quadratic form
$\hat h_\ell(\phi_\ell)=\hat{\bb a}^+B_\ell(\phi_\ell)\hat{\bb a}$, where
$\hat{\bb a}=(\hat a_1,\ldots,\hat a_N)^\trans$ collects the annihilation operators of $N$ bosonic modes, $\hat a_j^+$ creates a boson in mode $j$, $B_\ell$ is an $N\times N$ coefficient matrix, and $[\hat a_i,\hat a_j^+]=\delta_{ij}$. The bosonic Gaussian trace gives~\cite{Rubenstein2012PRA}
\begin{equation}
    W(\phi)=
    \det\!\left(I_N-\mc B(\phi)\right)^{-1}.
    \label{eq:trace_formula_number}
\end{equation}
Here $\mc B(\phi)=\ee^{B_1(\phi_1)}\ee^{B_2(\phi_2)}\cdots\ee^{B_L(\phi_L)}$ is the propagation coefficient matrix in the single-particle space.

If $\hat h_\ell$ contains pairing terms, particle number is not conserved and the one-body evolution is written in Nambu space. With
$\hat\chi=(\hat{\bb a},\hat{\bb a}^+)^\trans$, $[\hat\chi_\mu,\hat\chi_\nu]=\Omega_{\mu\nu}$, and
$\Omega=\left(\begin{smallmatrix}0&I_N\\-I_N&0\end{smallmatrix}\right)$, a quadratic slice is written as
$\hat h_\ell(\phi_\ell)=-\frac12\hat\chi^\trans\Omega A_\ell(\phi_\ell)\hat\chi+\xi_\ell(\phi_\ell)$, with
$\Omega A_\ell+A_\ell^\trans\Omega=0$. The scalar $\xi_\ell(\phi_\ell)$ includes the normal-ordering constant in this Nambu representation. The corresponding trace formula is
\begin{equation}
    W(\phi)=\exp\!\left[\sum_{\ell=1}^{L}\xi_\ell(\phi_\ell)\right]
    \bigl[(-1)^N\det(I_{2N}-\mc M(\phi))\bigr]^{-\frac12}.
    \label{eq:trace_formula_main}
\end{equation}
Here $\mc M(\phi)=\ee^{A_1(\phi_1)}\ee^{A_2(\phi_2)}\cdots\ee^{A_L(\phi_L)}$; Eq.~\eqref{eq:trace_formula_main} applies to Hamiltonians with explicit pairing terms.

BAFQMC performs importance sampling of auxiliary-field configurations using the weights in Eqs.~\eqref{eq:trace_formula_number} and~\eqref{eq:trace_formula_main}. Derivations of the trace formulas, fast-update algorithms, and observable estimators are given in Secs.~\ref{app:number_conserving_gaussian} and~\ref{app:paired_gaussian} of the SM~\cite{SM}.

\phantomsection\label{sec:positive}
\runinsection{Sign-problem-free Conditions}
 The central question is identifying the conditions that guarantee a positive weight for every auxiliary-field configuration in BAFQMC, thereby avoiding the sign problem. In fermionic AFQMC, the absence of the sign problem can often be established by analyzing the global symmetries and underlying mathematical structures of the HS-decoupled Hamiltonians, as developed extensively in Refs.~\cite{lang1993prc,wu2005prb,li2015prb,wang2015prl,wei2016prl,li2016prl,li2019arcmp,wei2024prb}. Inspired by this framework,  we propose two  sufficient criteria for bosonic models.

\begin{theorem}[TRS criterion]
\label{thm:trs}
 Assume that all $\hat h_\ell$ in \Eq{eq:Z_afqmc} share pairwise anticommuting time-reversal symmetries $\Theta_a$ with $\Theta_a^2=\pm I$ and $\Theta_a\Theta_b=-\Theta_b\Theta_a$ for $a\ne b$. They are sign-problem-free if they belong to the symmetry classes as shown in Table~\ref{tab:trs_periodic}.
\end{theorem}

For a particle-number-conserving $\hat h_\ell = \hat{\bb a}^+B_\ell\hat{\bb a}$ that preserves a TRS with $\Theta^2=-I$, every eigenmode of the full propagation matrix $\mc B$  in \Eq{eq:trace_formula_number} with eigenvalue $\lambda$ has a time-reversed partner with eigenvalue $\overline\lambda$~\cite{wu2005prb}. Furthermore, these partner modes remain linearly independent even when $\lambda$ is real. Consequently, each pair contributes a factor of $(1-\lambda)(1-\overline\lambda)=|1-\lambda|^2$ to the determinant in \Eq{eq:trace_formula_number}, yielding a strictly positive Boltzmann weight $W=\prod_{\rm pairs}|1-\lambda|^{-2}>0$.

Without the requirement of particle-number conservation, we can classify generic operators $\hat{h}_\ell$ according to their maximal set of mutually anticommuting TRSs, $\{\Theta_1,\cdots,\Theta_n\}$ (where $\Theta_a\Theta_b=-\Theta_b\Theta_a$ for $a\ne b$). Note that in this context, a TRS refers to a general antiunitary symmetry $\Theta = UK$, where $U$ is a unitary transformation and $K$ is complex conjugation. Based on these symmetry groups, we rigorously identify three fundamental sign-problem-free classes: $\{-,-\}$, $\{+,-\}$, and $\{+,+,+\}$, where each sign indicates whether $\Theta_a^2=+I$ or $-I$. The mechanisms guaranteeing the absence of the sign problem in these classes are detailed below.

\emph{\underline{Kramers class}:} For $\Theta_1^2=\Theta_2^2=-I$, the product $Q=\Theta_1\Theta_2$ obeys $Q^2=-I$ and generates a conserved quadratic charge, $\hat Q=-\hat\chi^\trans\Omega(\ii Q)\hat\chi/2$, with $[\hat Q,\hat h_\ell]=0$. Charge conservation separates the $Q=\pm\ii$ operator sectors, while $\Theta_1$ acts within each sector. Thus its propagation eigenvalues retain Kramers pairing, $\lambda,\overline\lambda$, and contribute $|1-\lambda|^2$ to the reduced determinant. Because a conserved charge exists in this class, the absence of the sign problem originates from the Kramers degeneracy. We therefore call this class the ``Kramers class''.

\emph{\underline{Pairing class}:} For the symmetry class $\{+,-\}$ with $\Theta_1^2=+I$ and $\Theta_2^2=-I$, the product $P=\Theta_1\Theta_2$ satisfies $P^2=I$ and separates the system into two independent bosonic subsystems characterized by eigenvalues $P=\pm 1$. Accordingly, the operator decomposes as $\hat{h}_\ell = \hat{h}^+_\ell + \hat{h}^-_\ell$, where $\hat{h}^{\pm}_\ell$ acts exclusively on the $P=\pm 1$ subspace. Because $P$ and $\Theta_1$ anticommute ($P\Theta_1=-\Theta_1P$), the antiunitary symmetry $\Theta_1$ exchanges these two subsystems and relates their evolution via complex conjugation. Their respective weights, $W_\pm$, therefore satisfy $W_-=\overline{W_+}$, yielding a non-negative total weight $W=W_+W_-=|W_+|^2\ge0$.

\emph{\underline{Split-orthogonal class}:} For the symmetry class with $\Theta_1^2=\Theta_2^2=\Theta_3^2=+I$, the first two symmetries define $Q=\Theta_1\Theta_2$ satisfying $Q^2=-I$. Projecting onto the $Q=+\ii$ sector, $\Theta_3$ pairs each annihilation direction with a creation direction, so the sector contains $N/2$ of each and the total Boltzmann weight takes the form $W\propto(-1)^{N/2}\det(I_N-\mc M_{+\ii})^{-1}$. Furthermore, it imposes the metric invariance $\mc M_{+\ii}^\trans\eta\mc M_{+\ii}=\eta$, where $\eta=\diag(I_{N/2},-I_{N/2})$ in an appropriate real basis, placing $\mc M_{+\ii}$ in the split-orthogonal group~\cite{wang2015prl}. Crucially, this structure fixes the sign of $\det(I_N-\mc M_{+\ii})$ to be $(-1)^{N/2}$, which cancels the factor from the $N/2$ bosonic creation operators and yields the strictly positive weight $W>0$.

 For symmetry classes with $n=3$ (other than $\{+,+,+\}$) and all classes with $n \geq 4$, where $n$ denotes the number of mutually anticommuting TRSs, the system necessarily possesses symmetries belonging to at least one of the $\{+,-\}$, $\{-,-\}$, or $\{+,+,+\}$ classes. Thus, they are inherently free from the sign problem. However, the remaining $\{+,+\}$ and $\{-\}$ classes can admit negative weights, as demonstrated by explicit counterexamples. Consequently, we rigorously prove that, based solely on symmetry requirements, there are exactly three fundamental sign-problem-free symmetry classes, as summarized in Table~\ref{tab:trs_periodic}. The Kramers and pairing classes are the bosonic counterparts of the two sign-free classes found in fermionic systems~\cite{li2016prl}. Intriguingly, the split-orthogonal class is sign-free only for bosons, possessing no fermionic counterpart. Detailed positivity proofs for these three fundamental classes are provided in Sec.~\ref{app:trs_proof} of the SM~\cite{SM}.

\begin{table}[t]
    \caption{\textbf{Periodic table of time-reversal symmetries for sign-problem-free bosonic AFQMC.} The Kramers class $\{-,-\}$, pairing class $\{+,-\}$, and split-orthogonal class $\{+,+,+\}$ are the three fundamental sign-problem-free classes, highlighted in bold blue. Each sign denotes $\Theta_a^2=\pm I$ for pairwise anticommuting time-reversal symmetries shared by the quadratic Hamiltonians at every imaginary-time slice. The label \emph{All} includes every possible sign assignment. Conversely, classes lacking these protective symmetries suffer from the sign problem and can yield negative Boltzmann weights.}
    \label{tab:trs_periodic}
    \vspace{6pt}
    \fontencoding{T1}\sffamily
    \begin{tikzpicture}[x=\columnwidth,y=1pt,
        every node/.style={inner sep=0pt,outer sep=0pt,font=\sffamily\footnotesize},
        heading/.style={text=white,font=\sffamily\bfseries\footnotesize,align=center},
        fundamental/.style={text=TRSAccent,font=\sffamily\bfseries\footnotesize}]
        \begin{scope}
            \clip[rounded corners=2pt] (0,0) rectangle (1,-120);
            \fill[TRSHeader]   (0,0)     rectangle (1,-22);
            \fill[TRSRowLight] (0,-22)   rectangle (1,-36.5);
            \fill[TRSRowShade] (0,-36.5) rectangle (1,-51);
            \fill[TRSRowLight] (0,-51)   rectangle (1,-78.5);
            \fill[TRSRowShade] (0,-78.5) rectangle (1,-105.5);
            \fill[TRSRowLight] (0,-105.5) rectangle (1,-120);
            \node[heading] at (0.075,-11) {No.\ of\\TRS};
            \node[heading] at (0.41,-11) {Sign-problem-free};
            \node[heading] at (0.835,-11) {Sign-problematic};
            \node at (0.075,-29.25) {0};
            \node at (0.41,-29.25) {None};
            \node at (0.835,-29.25) {No TRS};
            \node at (0.075,-43.75) {1};
            \node at (0.41,-43.75) {None};
            \node at (0.835,-43.75) {\{+\}, \{\textminus\}};
            \node at (0.075,-64.75) {2};
            \node[fundamental] at (0.41,-59) {\{\textminus,\textminus\} = Kramers class};
            \node[fundamental] at (0.41,-71) {\{+,\textminus\} = pairing class};
            \node at (0.835,-64.75) {\{+,+\}};
            \node at (0.075,-92) {3};
            \node[fundamental] at (0.41,-86) {\{+,+,+\} = split-orthogonal class};
            \node at (0.41,-98) {\{+,+,\textminus\}, \{+,\textminus,\textminus\}, \{\textminus,\textminus,\textminus\}};
            \node at (0.835,-92) {None};
            \node at (0.075,-112.75) {\(\geq\)\,4};
            \node at (0.41,-112.75) {All};
            \node at (0.835,-112.75) {None};
        \end{scope}
    \end{tikzpicture}
\end{table}

\begin{figure*}[t]
    \centering
    \includegraphics[width=\linewidth]{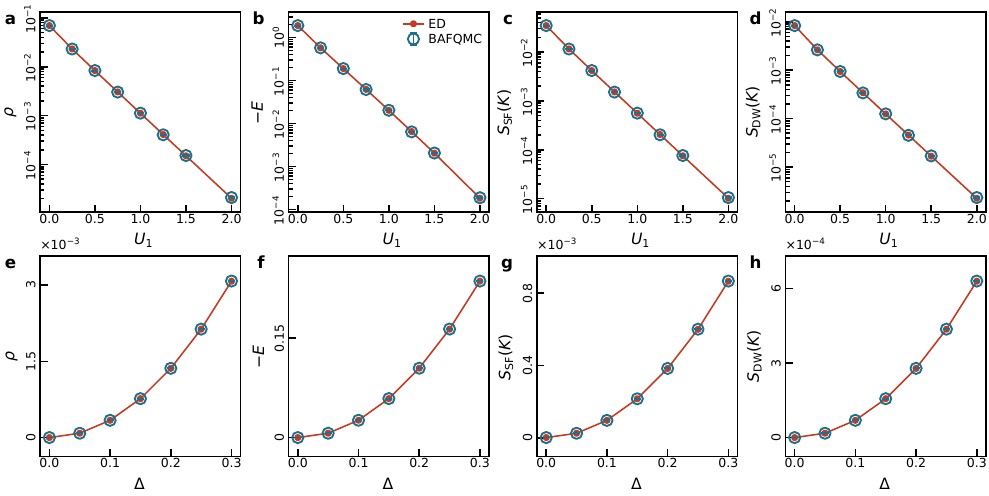}
    \caption{\textbf{BAFQMC versus exact diagonalization (ED) on a frustrated triangular lattice.} Results for Eq.~\eqref{eq:benchmark_pairing} on a periodic $3\times3$ lattice with $t=1$ and inverse temperature $\beta=4$. The simulations use $\hat H-\mu\sum_i(\hat n_{b,i}+\hat n_{c,i})$, where $\mu$ is the chemical potential. (a--d) Interaction scan at $\Delta=0$ and $\mu=-3.5$. (e--h) Pairing scan at $U_1=1$ and $\mu=-5$. The columns show the density per site $\rho$, negative energy $-E=-\langle\hat H\rangle$, and superfluid and density-wave structure factors $S_{\rm SF}(K)$ and $S_{\rm DW}(K)$ at a Brillouin-zone corner $K$. Red points are ED results; blue open circles are BAFQMC means with standard-error bars.}
    \label{fig:benchmark}
\end{figure*}

The TRS criteria guarantee positive global weights through symmetry-group constraints, irrespective of local frustration. In addition to these symmetry groups, RP~\cite{osterwalder1973axioms,lieb1989prl,wei2016prl} is another global mathematical structure that guarantees that the Boltzmann weights are non-negative. This property provides an analogous mechanism for bosonic systems, ultimately leading to our second criterion.

\begin{theorem}[RP criterion]
\label{thm:rp}
Let $\theta$ be the antilinear reflection defined by $\theta(\ii)=-\ii$, $\theta(\hat b_i)=\eta_i\hat c_i$, and $\theta(\hat b_i^+)=\overline{\eta_i}\hat c_i^+$, with phase factors $|\eta_i|=1$. If each $\hat h_\ell$ has the form
\begin{equation}
    \hat h_\ell =
    \hat q_\ell+\theta(\hat q_\ell)
    +\sum_{\mu,\nu} C_{\ell,\mu\nu}\,
    \hat X_{\ell\mu}\,\theta(\hat X_{\ell\nu}),
\label{eq:rp_generator}
\end{equation}
where $\hat q_\ell$ and $\hat X_{\ell\mu}$ are constructed exclusively from the $b$-boson operators and $C_\ell\succeq0$, then the weight $W=\Tr\prod_{\ell=1}^{L}\ee^{\hat h_\ell}$ is strictly positive whenever it is finite.
\end{theorem}
The detailed proof of Theorem~\ref{thm:rp} is given in Sec.~\ref{app:rp_proof} of the SM~\cite{SM}. For Gaussian Hamiltonian, Eq.~\eqref{eq:rp_generator} becomes a simple block test. Choosing the phase convention $\eta_i=1$ and grouping each flavor with its creation operators as $\hat\chi=(\hat{\bb b},\hat{\bb b}^+,\hat{\bb c},\hat{\bb c}^+)^\trans$, suppose every slice can be written as
\begin{equation}
    \hat h_\ell = \hat\chi^\trans
    \begin{pmatrix}
        D_\ell & F_\ell \\
        F_\ell^\trans & \overline{D_\ell}
    \end{pmatrix}
    \hat\chi + \xi_\ell,
    \qquad
    F_\ell=F_\ell^\hc\succeq0,\quad \xi_\ell\in\mathbb R .
    \label{eq:rp_block_form}
\end{equation}
Then the complete BAFQMC configuration weight is positive. The matrices $D_\ell$ and $\overline{D_\ell}$ describe quadratic terms within the $b$ and $c$ flavors, respectively, while $F_\ell$ couples the two flavors, allowing RP to protect positivity beyond symmetry-group constraints.

Together, TRS and RP define the sign-free region shown schematically in Fig.~\ref{fig:schematic}: these conditions guarantee positive global auxiliary-field weights even on frustrated lattices.

\phantomsection\label{sec:examples}
\runinsection{Examples and applications}
A crucial frustrated model identified in our framework is a two-flavor Bose-Hubbard model on the triangular lattice. With $\hat n_{b,i}=\hat b_i^+\hat b_i$ and $\hat n_{c,i}=\hat c_i^+\hat c_i$, consider
\begin{equation}
\begin{aligned}
    \hat H ={}&
    t\sum_{\langle ij\rangle}
    \left(\hat b_i^+\hat b_j+\hat c_i^+\hat c_j+\mathrm{h.c.}\right)
    -\sum_i\left(\Delta\,\hat b_i^+\hat c_i^+ +\mathrm{h.c.}\right)\\
    &+U_1\sum_i\left(\hat n_{b,i}-\hat n_{c,i}\right)^2
    +U_2\sum_i\left(\hat n_{b,i}+\hat n_{c,i}\right)^2,
\end{aligned}
\label{eq:benchmark_pairing}
\end{equation}
Here $\langle ij\rangle$ denotes nearest-neighbor bonds, $t$ is the hopping amplitude, $\Delta$ is the pairing amplitude, and $U_1,U_2$ are the Hubbard interaction strengths. For $t>0$, the hopping is frustrated and produces a sign problem in world-line QMC and SSE. However, BAFQMC avoids this obstruction. For $U_1\ge0$ and $U_2\le0$, the HS transformation in Sec.~\ref{app:hs_general} of the SM~\cite{SM} couples an imaginary field to $\hat n_{b,i}-\hat n_{c,i}$ and a real field to $\hat n_{b,i}+\hat n_{c,i}$, giving the two flavors complex-conjugate potentials. At $\Delta=0$, every slice preserves the TRS $\Theta=\ii\sigma_y K$ in flavor space, with $\Theta^2=-I$, and belongs to the sign-problem-free Kramers class. At $\Delta\ne0$, pairing yields the positive coupling block $F\propto|\Delta|I\succeq0$ in Eq.~\eqref{eq:rp_block_form}, and the RP criterion ensures sign-problem-free simulations for $U_2=0$~\cite{SM}.

Figure~\ref{fig:benchmark} compares BAFQMC with finite-temperature exact diagonalization (ED) on a $3\times3$ triangular lattice with periodic boundary conditions. We set $U_2=0$ and scan $U_1$ at $\Delta=0$, then scan $\Delta$ at $U_1=1$. The BAFQMC results coincide with ED within statistical errors for the density, energy, and both structure factors. Similar agreement for the attractive $U_2$ interaction is shown in Sec.~\ref{app:benchmark_details} of the SM~\cite{SM}. These benchmarks establish the reliability of BAFQMC.

Another important application of our symmetry criterion is the study of odd-parity bosonic pairing systems. Bose exchange symmetry gives the antisymmetric pair $\hat b_i^+\hat c_j^+-\hat b_j^+\hat c_i^+$. We prove in the SM~\cite{SM} that the split-orthogonal class is the unique fundamental TRS class guaranteeing sign-free odd-parity bosonic pairing, including $p$-wave pairing. This flavor-singlet pairing channel is central to finite-momentum superfluidity in $p$-wave resonant Bose gases~\cite{radzihovskyChoi2009pwave} and to the singlet bonds in Schwinger-boson theories of quantum spin liquids~\cite{readSachdev1991frustrated,sachdev1992kagome}. BAFQMC therefore opens unbiased access to broad classes of frustrated bosonic and spin models of longstanding physical interest.

We reveal more sign-problem-free models in the SM~\cite{SM}, including interacting $p$-orbital bosons and pairing-induced chiral bands. These models provide theoretical platforms for studying unconventional Bose-Einstein condensates~\cite{liuWu2006orbital,wu2009unconventional,wuLiuMoore2006stripe,liYuanHemmerichLi2018prl,wangLuoLiu2023stripe} and topological bosonic matter~\cite{peanoHoudeBrendel2016squeezing,bardynKarzigRefaelLiew2016bogoliubov,peanoHoudeMarquardt2016topological,slimWanjuraBrunelli2024bosonKitaev}.

\phantomsection\label{sec:concluding}
\runinsection{Concluding remarks}
In summary, we have formulated a BAFQMC framework that establishes unifying principles for tackling the sign problem in interacting bosonic systems. Our two sufficient criteria, rooted in time-reversal symmetry and reflection positivity, provide practical guidance for guaranteeing sign-problem-free simulation. We validated this approach by simulating a frustrated two-flavor Bose-Hubbard model on the triangular lattice, achieving excellent agreement with exact diagonalization within statistical errors. Crucially, our results demonstrate a fundamental shift in the organizing principle for bosonic sign-free simulations: from the positivity of local matrix elements to the global symmetries and mathematical structures of the Hamiltonian. As depicted in Fig.~\ref{fig:schematic}, this unlocks a new sign-problem-free route into the regime of geometric frustration, transcending the natural boundaries of world-line QMC and SSE.

This result also connects the bosonic and fermionic auxiliary-field viewpoints. In both cases, symmetry conditions on the HS-decoupled one-body problem can define sign-free regions. The statistical distinction appears in the Gaussian trace: fermions produce $\det[I_N+\mc B(\phi)]$, whereas bosons produce the inverse determinant $\det[I_N-\mc B(\phi)]^{-1}$. The inverse determinant can generate large weights near its poles~\cite{Rubenstein2012PRA}, giving a bosonic analogue of the infinite-variance problem known in fermionic AFQMC~\cite{Shi2016PRE,alexandru2023prd,wan2025infinite}. This numerical difficulty is most relevant when the sampled trajectory approaches such poles, for example at higher chemical potential, and its severity depends on the Hamiltonian, the HS decomposition, and the determinant representation.

A major frontier opened by this work is the ability to revisit frustrated bosonic and spin systems from a new perspective. Frustration has long obstructed local-weight QMC~\cite{henelius2000prb}, leaving the study of frustrated Bose-Hubbard models, frustrated quantum magnets, and spin-liquid physics largely reliant on methods outside the realm of sign-free QMC~\cite{wang2009prl,zaletel2014prb,yan2011science,zhu2015prb,zhou2017rmp}. Whenever such models admit an RP or TRS HS decoupling, BAFQMC provides a constructive route to sign-problem-free simulation. Similar positivity principles should also guide bosonic projector AFQMC, where ground-state properties can be targeted directly~\cite{purwanto2004pre,Wei2025arxiv}. Canonical fixed-particle-number formulations are another natural direction~\cite{Shen2020JCP,Shen2023PRE}, since removing grand-canonical high-occupation sectors offers a practical way to avoid determinant-pole variance problems while retaining the sign-free condition.

\runinsection{Acknowledgments}
We thank Lei Wang for helpful discussions. Y.-K.Y. and Z.-X.L. are supported by the National Natural Science Foundation of China under Grant Nos. 12347107 and 12474146, and Beijing Natural Science Foundation under Grant No. JR25007. Z.-X.L. is supported by the New Cornerstone Investigator Program.

\runinsection{Data availability}
The benchmark data and scripts for reproducing the numerical results are available in the BAFQMC repository~\cite{bafqmc_code}.

\runinsection{Code availability}
The BAFQMC and ED codes used in this work are available under the MIT license in Ref.~\cite{bafqmc_code}.

\pdfbookmark[1]{References}{main.references}
\bibliography{references}

\clearpage
\onecolumngrid
\newpage
\widetext
\thispagestyle{empty}

\makeatletter
\let\addcontentsline\oldaddcontentsline
\makeatother

\setcounter{section}{0}
\setcounter{subsection}{0}
\setcounter{equation}{0}
\setcounter{figure}{0}
\setcounter{table}{0}
\setcounter{secnumdepth}{2}
\renewcommand{\thesection}{\Roman{section}}
\renewcommand{\thesubsection}{\Alph{subsection}}
\renewcommand{\theequation}{S\arabic{equation}}
\renewcommand{\thefigure}{S\arabic{figure}}
\renewcommand{\thetable}{S\arabic{table}}
\renewcommand{\theHsection}{SM.\Roman{section}}
\renewcommand{\theHsubsection}{SM.\Roman{section}.\Alph{subsection}}
\renewcommand{\theHequation}{SM.\arabic{equation}}
\renewcommand{\theHfigure}{SM.\arabic{figure}}
\renewcommand{\theHtable}{SM.\arabic{table}}

\renewcommand\floatpagefraction{0.9}
\renewcommand\textfraction{0.1}

\pdfbookmark[0]{Supplemental Material}{SM}
\begin{center}
    \vspace{3em}
    {\Large\textbf{Supplemental Material for}}\\
    \vspace{1em}
    {\large\textbf{Sign-problem-free Bosonic Auxiliary Field Quantum Monte Carlo}}\\
    \vspace{0.5em}
\end{center}

\tableofcontents

\section{Hubbard-Stratonovich transformation}
\label{app:hs_general}

The HS transformation expresses the interacting Boltzmann factor as an integral over quadratic bosonic propagators coupled to auxiliary fields. Consider
\begin{equation}
    \hat H=\hat H_0+\hat H_{\mathrm{int}},\qquad
    \hat H_{\mathrm{int}}=\sum_\gamma\lambda_\gamma\hat O_\gamma^2,
\end{equation}
where $\hat H_0$ is quadratic and the Hermitian one-body operator $\hat O_\gamma$ specifies interaction channel $\gamma$. Divide the imaginary-time interval into $L_\tau$ slices of width $\Delta\tau=\beta/L_\tau$, indexed by $\ell$. A first-order Trotter-Suzuki decomposition gives
\begin{equation}
    Z=\Tr\,\ee^{-\beta\hat H}
    =\Tr\left[\prod_{\ell=1}^{L_\tau}\ee^{-\Delta\tau\hat H_0}
    \prod_\gamma\ee^{-\Delta\tau\lambda_\gamma\hat O_\gamma^2}\right]+O(\Delta\tau).
\end{equation}
Each interaction factor can now be decoupled exactly. For an attractive channel, $\lambda_\gamma<0$, the Gaussian identity gives
\begin{equation}
    \ee^{-\Delta\tau\lambda_\gamma\hat O_\gamma^2}
    =
    \int\frac{\dd\phi_{\gamma\ell}}{\sqrt{2\pi}}\,
    \ee^{-\phi_{\gamma\ell}^{2}/2
    +\sqrt{-2\Delta\tau\lambda_\gamma}\,\phi_{\gamma\ell}\hat O_\gamma}.
    \label{eq:hs_attractive}
\end{equation}
For a repulsive channel, $\lambda_\gamma>0$, a real Gaussian field instead couples to $\hat O_\gamma$ with an imaginary coefficient,
\begin{equation}
    \ee^{-\Delta\tau\lambda_\gamma\hat O_\gamma^2}
    =
    \int\frac{\dd\phi_{\gamma\ell}}{\sqrt{2\pi}}\,
    \ee^{-\phi_{\gamma\ell}^{2}/2
    +\ii\sqrt{2\Delta\tau\lambda_\gamma}\,\phi_{\gamma\ell}\hat O_\gamma}.
    \label{eq:hs_repulsive}
\end{equation}
In both channels, the interaction is replaced by a one-body coupling to the real field $\phi_{\gamma\ell}$. At fixed fields, the decoupled interaction factors and $\ee^{-\Delta\tau\hat H_0}$ form the quadratic evolution whose bosonic trace is $W(\phi)$. Integrating this weight over all fields with the product Gaussian measure $P_{\rm HS}(\phi)$ gives the partition function in Eq.~\eqref{eq:Z_afqmc}. Here $\phi$ denotes the complete auxiliary-field configuration.

For the two-flavor algorithms, we include both a repulsive relative-density interaction $U_1(\hat n_{b,i}-\hat n_{c,i})^2$ and an attractive total-density interaction $U_2(\hat n_{b,i}+\hat n_{c,i})^2$, with $U_1\ge0$ and $U_2\le0$. The main-text model in Eq.~\eqref{eq:benchmark_pairing} focuses on $U_1\ge0$ with $U_2=0$. Denoting the real auxiliary fields for the total- and relative-density channels by $\phi_{i\ell}$ and $\varphi_{i\ell}$, respectively, Eqs.~\eqref{eq:hs_attractive} and~\eqref{eq:hs_repulsive} give
\begin{equation}
    \ee^{-\Delta\tau U_2(\hat n_{b,i}+\hat n_{c,i})^2}
    =
    \int \frac{\dd\phi_{i\ell}}{\sqrt{2\pi}}\,
    \ee^{-\phi_{i\ell}^2/2+\alpha_2\phi_{i\ell}(\hat n_{b,i}+\hat n_{c,i})},
    \qquad
    \alpha_2=\sqrt{-2\Delta\tau U_2},
\label{eq:benchmark_hs_total}
\end{equation}
and
\begin{equation}
    \ee^{-\Delta\tau U_1(\hat n_{b,i}-\hat n_{c,i})^2}
    =
    \int \frac{\dd\varphi_{i\ell}}{\sqrt{2\pi}}\,
    \ee^{-\varphi_{i\ell}^2/2+\ii\alpha_1\varphi_{i\ell}(\hat n_{b,i}-\hat n_{c,i})},
    \qquad
    \alpha_1=\sqrt{2\Delta\tau U_1}.
\label{eq:benchmark_hs_relative}
\end{equation}
At fixed fields, the onsite potentials are
\begin{equation}
    v_{b,i\ell}=\alpha_2\phi_{i\ell}+\ii\alpha_1\varphi_{i\ell},\qquad
    v_{c,i\ell}=\alpha_2\phi_{i\ell}-\ii\alpha_1\varphi_{i\ell}.
    \label{eq:benchmark_hs_potentials}
\end{equation}
These coefficients enter the dimensionless onsite generator $\sum_i(v_{b,i\ell}\hat n_{b,i}+v_{c,i\ell}\hat n_{c,i})$. The total-density field shifts the two flavors equally, while the relative-density field gives opposite imaginary shifts, ensuring $v_{c,i\ell}=\overline{v_{b,i\ell}}$. For grand-canonical sampling with $\hat H-\mu\hat N$, the chemical potential adds $+\Delta\tau\mu\sum_i(\hat n_{b,i}+\hat n_{c,i})$ to the generator.

\section{Number-conserving Gaussian formalism}
\label{app:number_conserving_gaussian}

For a fixed auxiliary-field configuration, the HS transformation reduces the interacting problem to bosons propagating in an external field. When this propagation conserves particle number, the sum over all boson occupations and the equal-time correlations can both be evaluated from an $N\times N$ single-particle matrix. These two quantities determine the configuration weight, the observable estimators, and the acceptance of local field updates.

\subsection{Trace formula}
\label{app:trace_number}

Let $\hat{\bb a}=(\hat a_1,\ldots,\hat a_N)^\trans$ collect the annihilation operators of the $N$ bosonic modes. At fixed auxiliary fields, the many-body propagator and its single-particle matrix are
\begin{equation}
    \hat U=\prod_{\ell=1}^{L}\exp[\hat{\bb a}^+B_\ell(\phi_\ell)\hat{\bb a}],
    \qquad
    \mc B=\ee^{B_1(\phi_1)}\cdots\ee^{B_L(\phi_L)}.
\end{equation}
Here $B_\ell$ is the coefficient matrix of the quadratic generator, $\ee^{B_\ell}$ propagates a single particle through that factor, and $\mc B$ describes the complete imaginary-time evolution. We suppress the field dependence below.

The bosonic trace sums over every occupation of the single-particle modes. If $\mc B$ is diagonal with eigenvalues $r_1,\ldots,r_N$, a state with $n_i$ bosons in mode $i$ acquires the factor $r_i^{n_i}$. Summing these factors gives
\begin{equation}
    \Tr\,\hat U
    =
    \sum_{n_1,\ldots,n_N\ge0}\prod_{i=1}^{N}r_i^{n_i}
    =
    \prod_{i=1}^{N}\frac{1}{1-r_i}=\det(I_N-\mc B)^{-1},
    \qquad |r_i|<1.
\end{equation}
Thus the inverse determinant arises from the unrestricted bosonic occupations, with each mode contributing a geometric series~\cite{Rubenstein2012PRA}.

In general, successive imaginary-time slices mix the single-particle modes, so their evolution need not be diagonal in the chosen basis. Coherent states provide a direct way to evaluate the trace in this case because a number-conserving quadratic propagator maps a coherent state to another coherent state, with linearly transformed amplitudes. We use the unnormalized states $|z\rangle=\exp(\hat{\bb a}^+z)|0\rangle$, where $z$ is a vector of complex amplitudes. Their overlap and resolution of the identity are
\begin{equation}
    \langle z|w\rangle=\ee^{z^\hc w},
    \qquad
    I=\int\frac{\dd^{2N}z}{\pi^N}\,\ee^{-z^\hc z}\,|z\rangle\langle z|.
\end{equation}
Each quadratic factor acts linearly on the creation operators,
\begin{equation}
    [\hat{\bb a}^+B_\ell\hat{\bb a},\hat a_i^+]
    =\sum_j\hat a_j^+(B_\ell)_{ji},
    \qquad
    \ee^{\hat{\bb a}^+B_\ell\hat{\bb a}}\hat{\bb a}^+
    \ee^{-\hat{\bb a}^+B_\ell\hat{\bb a}}
    =\hat{\bb a}^+\ee^{B_\ell}.
\end{equation}
Since the vacuum is invariant, each factor sends $|z\rangle$ to $|\ee^{B_\ell}z\rangle$, and the full evolution gives $\hat U|z\rangle=|\mc B z\rangle$. The Fock-space trace therefore becomes an integral over the coherent-state amplitudes,
\begin{equation}
    \Tr\,\hat U
    =
    \int\frac{\dd^{2N}z}{\pi^N}\,\ee^{-z^\hc z}
    \langle z|\hat U|z\rangle
    =
    \int\frac{\dd^{2N}z}{\pi^N}
    \exp[-z^\hc(I_N-\mc B)z].
\end{equation}
When the Hermitian part of $I_N-\mc B$ is positive definite, this Gaussian integral converges. The standard complex Gaussian integral then gives
\begin{equation}
    \Tr\,\hat U=\det(I_N-\mc B)^{-1}.
    \label{eq:number_trace_product}
\end{equation}
Thus the sum over bosonic occupations becomes a Gaussian integral over coherent-state amplitudes, giving the same inverse determinant for a general propagation matrix. This is Eq.~\eqref{eq:trace_formula_number}. For fixed-field configurations with divergent bare traces, Sec.~\ref{app:regularization} derives the same determinant expression as the finite part of a Gaussian-regulated occupation sum.

\subsection{Equal-time Green function}
\label{app:green_number}

The same propagation matrix determines the particle density and equal-time correlations in each auxiliary-field configuration. Define
\begin{equation}
    G_{ij}=\langle\hat a_i\hat a_j^+\rangle_\phi,
\end{equation}
where $\langle\hat O\rangle_\phi=\Tr(\hat O\hat U)/\Tr\hat U$. The bosonic commutator gives $\langle\hat a_j^+\hat a_i\rangle_\phi=G_{ij}-\delta_{ij}$. Thus $G_{ii}-1$ is the occupation of mode $i$ at fixed fields, while the off-diagonal elements describe one-body coherence.

The two operator orderings differ by the bosonic commutator. They are also related by moving an annihilation operator through the full imaginary-time evolution and using cyclicity of the trace. These two relations determine the Green function. To obtain the propagation relation, the commutator $[\hat{\bb a}^+B_\ell\hat{\bb a},\hat a_i]=-\sum_j(B_\ell)_{ij}\hat a_j$ implies
\begin{equation}
    \hat{\bb a}\,\ee^{\hat{\bb a}^+B_\ell\hat{\bb a}}
    =\ee^{\hat{\bb a}^+B_\ell\hat{\bb a}}\ee^{B_\ell}\hat{\bb a},
    \qquad
    \hat{\bb a}\hat U=\hat U\,\mc B\hat{\bb a}.
\end{equation}
Cyclicity of the trace then relates the two operator orderings,
\begin{equation}
    \langle\hat a_j^+\hat a_i\rangle_\phi
    =\frac{\Tr(\hat a_j^+\hat a_i\hat U)}{\Tr\hat U}
    =\frac{\Tr(\hat a_i\hat U\hat a_j^+)}{\Tr\hat U}
    =\sum_k\mc B_{ik}G_{kj}.
\end{equation}
Equating this result with $G_{ij}-\delta_{ij}$ gives $G-I_N=\mc B G$. Solving for $G$ yields
\begin{equation}
    G=(I_N-\mc B)^{-1},
    \qquad
    \langle\hat a_j^+\hat a_i\rangle_\phi=[\mc B(I_N-\mc B)^{-1}]_{ij}=G_{ij}-\delta_{ij},
\end{equation}
This is the number-conserving Green function used in the updates below. In the diagonal case, $\langle\hat n_i\rangle_\phi=r_i/(1-r_i)$ is the mean occupation obtained from the geometric series above. Higher equal-time correlators follow from Wick's theorem because $\hat U$ is Gaussian. The Green function also determines how the configuration weight changes under a local auxiliary-field update.

\subsection{Local updates and Metropolis ratio}
\label{app:update_number}

A Monte Carlo sweep proposes local changes of the auxiliary fields. Each proposal requires a weight ratio to decide acceptance and, if accepted, an updated Green function for subsequent proposals and measurements. Consider a local proposal $\phi_{i\ell}\to\phi_{i\ell}'$ that changes the onsite potential of mode $i$ at slice $\ell$. This multiplies the single-particle propagation amplitude on that mode by a factor $1+\delta$, whose dependence on the field change is specified by the HS transformation. In single-particle space, the change is therefore $I_N+\Delta_\ell$, with $\Delta_\ell=\delta\ket{i}\bra{i}$ and $\ket{i}$ the basis vector for that mode.

Cyclicity of the trace allows the changed onsite factor to be placed at the left end of the full product without changing the weight. With the slice boundaries chosen so that this factor is leftmost in $\ee^{B_\ell}$, define
\begin{equation}
    \mc B_\ell=\ee^{B_\ell}\ee^{B_{\ell+1}}\cdots\ee^{B_L}\ee^{B_1}\cdots\ee^{B_{\ell-1}},
    \qquad
    G_\ell=(I_N-\mc B_\ell)^{-1}.
\end{equation}
The matrix $\mc B_\ell$ contains the full imaginary-time evolution, ordered from slice $\ell$. The corresponding $G_\ell$ measures correlations at the imaginary-time position where this product begins, called the equal-time cut. All cyclic orderings give the same configuration weight.

\emph{Weight ratio.} At this cut, the onsite change gives $\ee^{B_\ell'}=(I_N+\Delta_\ell)\ee^{B_\ell}$ and hence $\mc B_\ell'=(I_N+\Delta_\ell)\mc B_\ell$. A field acting on several local modes has the same form, with nonzero matrix elements of $\Delta_\ell$ confined to those modes. Using $G_\ell\mc B_\ell=G_\ell-I_N$ gives
\begin{equation}
    I_N-\mc B_\ell'
    =\left[I_N+\Delta_\ell(I_N-G_\ell)\right](I_N-\mc B_\ell).
\end{equation}
Taking determinants yields the ratio of the bosonic weights,
\begin{equation}
    \frac{W'}{W}
    =\det\!\left[I_N+\Delta_\ell(I_N-G_\ell)\right]^{-1}.
    \label{eq:update_ratio_number}
\end{equation}
For the single-mode change $\Delta_\ell=\delta\ket{i}\bra{i}$, the matrix determinant lemma reduces the weight ratio to $W'/W=[1+\delta(1-(G_\ell)_{ii})]^{-1}$, which requires only the diagonal element $(G_\ell)_{ii}=\bra{i}G_\ell\ket{i}$ at the updated mode.

The acceptance probability includes the Gaussian HS measure as well as the bosonic weight. For a symmetric proposal $\phi_{i\ell}\to\phi_{i\ell}'$ in a sign-problem-free calculation,
\begin{equation}
    R=\exp\left[-\frac12(\phi_{i\ell}'^{\,2}-\phi_{i\ell}^{2})\right]
    \det\!\left[I_N+\Delta_\ell(I_N-G_\ell)\right]^{-1},
    \qquad
    p(\phi_{i\ell}\to\phi_{i\ell}')=\min(1,R).
\end{equation}
The exponential is the ratio of the Gaussian weights for the field being updated. When a field acts on several flavors, $W'/W$ includes all of their contributions.

\emph{Green-function update.} Once a proposal is accepted, the preceding factorization also determines the new correlations. Defining $C_\ell=I_N+\Delta_\ell(I_N-G_\ell)$ gives
\begin{equation}
    G_\ell'=(I_N-\mc B_\ell')^{-1}=G_\ell C_\ell^{-1}.
    \label{eq:number_green_update_general}
\end{equation}
For the onsite change, $C_\ell=I_N+\delta\ket{i}\bra{i}(I_N-G_\ell)$. The Sherman-Morrison identity yields
\begin{equation}
    C_\ell^{-1}
    =I_N-\frac{\delta\ket{i}(\bra{i}-\bra{i}G_\ell)}
                    {1+\delta(1-(G_\ell)_{ii})},
    \label{eq:number_sherman_morrison}
\end{equation}
and hence
\begin{equation}
    G_\ell'
    =G_\ell-\frac{\delta\,G_\ell\ket{i}(\bra{i}-\bra{i}G_\ell)}
                       {1+\delta(1-(G_\ell)_{ii})}.
\end{equation}
The local potential change modifies correlations throughout the system through the column $G_\ell\ket{i}$ and the row $\bra{i}(I_N-G_\ell)$, which connect the other modes to the updated mode $i$. Their outer product gives a rank-one correction to the full Green function. For a change acting on several local modes, the determinant lemma and the Woodbury identity reduce the calculation to that local subspace and its correlations with the rest of the system.

\emph{Two-flavor realization.} The two-flavor algorithm~\cite{bafqmc_code} includes the total- and relative-density channels of Sec.~\ref{app:hs_general}. The number-conserving main-text model sets $\Delta=0$ and $U_2=0$. The two flavors experience complex-conjugate HS potentials, and the quadratic terms entering each slice are
\begin{equation}
\begin{aligned}
    \hat h_\ell^{(0)}
    ={}&\sum_{ij}\hat b_i^+K_{ij}\hat b_j
       +\sum_i v_{b,i\ell}\hat n_{b,i}\\
       &+\sum_{ij}\hat c_i^+K_{ij}\hat c_j
       +\sum_i v_{c,i\ell}\hat n_{c,i},
\end{aligned}
    \label{eq:number_benchmark_slice}
\end{equation}
where the real matrix $K$ contains the hopping and chemical-potential contributions, including their imaginary-time step. The potentials $v_{b,i\ell}=\alpha_2\phi_{i\ell}+\ii\alpha_1\varphi_{i\ell}$ and $v_{c,i\ell}=\overline{v_{b,i\ell}}$ follow from Eqs.~\eqref{eq:benchmark_hs_total} and~\eqref{eq:benchmark_hs_relative}.

On a lattice of $N_s$ sites, the two flavors contain $N=2N_s$ modes, and each flavor propagates in an $N_s$-dimensional single-particle space. Let $\ee^{B_{b,\ell}}$ and $\ee^{B_{c,\ell}}$ denote the ordered products of kinetic and onsite propagators on one slice. For local updates, choose the equal-time cut immediately before the onsite factor, so that its change acts by left multiplication. The cyclic products are
\begin{equation}
    \mc B_{\sigma,\ell}
    =\ee^{B_{\sigma,\ell}}\ee^{B_{\sigma,\ell+1}}\cdots\ee^{B_{\sigma,L}}\ee^{B_{\sigma,1}}\cdots\ee^{B_{\sigma,\ell-1}},
    \qquad \sigma=b,c,
    \label{eq:number_cyclic_products_benchmark}
\end{equation}
with Green functions
\begin{equation}
    G_{b,\ell}=(I_{N_s}-\mc B_{b,\ell})^{-1},
    \qquad
    G_{c,\ell}=(I_{N_s}-\mc B_{c,\ell})^{-1}.
    \label{eq:number_green_bc_def}
\end{equation}
The real hopping and conjugate onsite potentials imply $\ee^{B_{c,\ell}}=\overline{\ee^{B_{b,\ell}}}$ at every slice. Consequently,
\begin{equation}
    \mc B_{c,\ell}=\overline{\mc B_{b,\ell}},
    \qquad
    G_{c,\ell}=\overline{G_{b,\ell}}.
    \label{eq:conj_green_benchmark}
\end{equation}
It is therefore sufficient to evaluate and stabilize the $b$-flavor propagation matrix and Green function. The corresponding $c$-flavor quantities follow by complex conjugation.

A local field proposal multiplies the $b$-flavor onsite propagator by $1+\delta_b$. For a total-density field change, $\delta_b=\ee^{\alpha_2(\phi_{i\ell}'-\phi_{i\ell})}-1$. For a relative-density field change, $\delta_b=\ee^{\ii\alpha_1(\varphi_{i\ell}'-\varphi_{i\ell})}-1$. The resulting single-flavor weight ratio is
\begin{equation}
    r_b=\left[1+\delta_b(1-(G_{b,\ell})_{ii})\right]^{-1}.
    \label{eq:single_flavor_local_ratio}
\end{equation}
Since $r_c=\overline{r_b}$, the full two-flavor ratio is $W'/W=r_b r_c=|r_b|^2$. Combining it with the Gaussian HS ratio gives the Metropolis ratio and acceptance probability used for the $\Delta=0$ benchmark,
\begin{equation}
    R=\exp\left[-\frac12(x_{i\ell}'^{\,2}-x_{i\ell}^{2})\right]|r_b|^2,
    \qquad
    p(x_{i\ell}\to x_{i\ell}')=\min(1,R),
    \label{eq:benchmark_number_local_ratio}
\end{equation}
where $x_{i\ell}$ denotes the real HS field being updated, either $\phi_{i\ell}$ or $\varphi_{i\ell}$. After acceptance, the Green function becomes
\begin{equation}
    G_{b,\ell}'=G_{b,\ell}
    -\delta_b r_b\,G_{b,\ell}\ket{i}
    \left(\bra{i}-\bra{i}G_{b,\ell}\right).
    \label{eq:number_green_rank_one}
\end{equation}
The $c$-flavor update is the complex conjugate of Eq.~\eqref{eq:number_green_rank_one}.

\subsection{Numerical stabilization}
\label{app:stabilization_number}

Different single-particle components grow or decay at different rates under imaginary-time evolution. At low temperature, the long propagation can make their amplitudes differ by many orders of magnitude. Direct matrix multiplication then loses information about the smaller components through roundoff error, compromising the Green function $(I_{N_s}-\mc B_{b,\ell})^{-1}$. We retain this information by periodically factorizing the propagation matrix and storing its large and small scales separately, following the stabilization strategy used in determinantal QMC~\cite{assaadEvertz2008}.

During a sweep, accepted proposals update $G_{b,\ell}$ through Eq.~\eqref{eq:number_green_rank_one}. Advancing the equal-time cut to slice $\ell+1$ cyclically permutes the product in Eq.~\eqref{eq:number_cyclic_products_benchmark},
\begin{equation}
    \mc B_{b,\ell+1}=\ee^{-B_{b,\ell}}\mc B_{b,\ell}\ee^{B_{b,\ell}}.
\end{equation}
The Green function transforms accordingly,
\begin{equation}
    G_{b,\ell+1}=\ee^{-B_{b,\ell}}G_{b,\ell}\ee^{B_{b,\ell}}.
    \label{eq:number_wrap_similarity}
\end{equation}
These updates propagate the correlations between nearby slices. To control accumulated roundoff error, the sweep is divided into windows of $N_{\rm wrap}$ slices, with $N_{\rm wrap}=10$ in the reported benchmarks. At each window boundary, $G_{b,\ell}$ is recomputed from the stabilized full cyclic product.

The factorization records the scales accumulated during propagation in a diagonal matrix $D$, while the remaining factors describe the mixing of single-particle components. To construct it, consider a short window beginning at slice $j$, with the same ordering as Eq.~\eqref{eq:number_cyclic_products_benchmark},
\begin{equation}
    X_j=\ee^{B_{b,j}}\ee^{B_{b,j+1}}\cdots\ee^{B_{b,j+N_{\rm wrap}-1}},
\end{equation}
with indices understood modulo $L$. A column-pivoted QR factorization, $X_jP_j=Q_jR_j$, gives a unitary $Q_j$, an upper-triangular $R_j$, and a permutation matrix $P_j$. Extracting the diagonal magnitudes $(D_j)_{ii}=|(R_j)_{ii}|$ gives
\begin{equation}
    X_j=U_jD_jV_j,
    \qquad U_j=Q_j,
    \qquad V_j=D_j^{-1}R_jP_j^\trans.
    \label{eq:window_udv_number}
\end{equation}
Here $U_j$ is unitary, while $V_j$ is a normalized upper-triangular factor followed by a permutation. To accumulate a new window on the left of a stored product $UDV$, first multiply the window into $U$ and scale the resulting columns by $D$. Factor this matrix by pivoted QR and extract its diagonal magnitudes to obtain the new $U$ and $D$. Absorb the remaining triangular factor and permutation into the stored $V$. Repeating this procedure after each window keeps the propagation scales explicit throughout the accumulation.

To reconstruct the Green function at slice $\ell$, combine the two intervals that make up its cyclic product: slices $\ell,\ldots,L$ and slices $1,\ldots,\ell-1$. Their factorizations are
\begin{equation}
    \ee^{B_{b,\ell}}\cdots\ee^{B_{b,L}}=U_RD_RV_R,
    \qquad
    \ee^{B_{b,1}}\cdots\ee^{B_{b,\ell-1}}=V_LD_LU_L.
\end{equation}
The second ordering follows by factorizing the Hermitian conjugate of that interval's product and then taking the Hermitian conjugate back. Both $U_R$ and $U_L$ are unitary. With this choice, the complete propagation has unitary factors at its outer ends,
\begin{equation}
    \mc B_{b,\ell}=U_RD_RV_RV_LD_LU_L.
    \label{eq:number_stab_product}
\end{equation}
The outer unitary factors can be removed by their Hermitian conjugates, leaving a middle matrix formed directly from the stored $V_R$, $V_L$, and diagonal scales. Explicitly,
\begin{equation}
    I_{N_s}-\mc B_{b,\ell}
    =U_R\left[U_R^\hc U_L^\hc-D_RV_RV_LD_L\right]U_L.
\end{equation}
Define the matrix between the two unitary factors by
\begin{equation}
    M_{\rm stab}=U_R^\hc U_L^\hc-D_RV_RV_LD_L.
    \label{eq:number_mstab_def}
\end{equation}
The reconstructed Green function is therefore
\begin{equation}
    G_{b,\ell}=U_L^\hc M_{\rm stab}^{-1}U_R^\hc.
    \label{eq:number_stab_green}
\end{equation}
The interval factorizations preserve the information in the long propagation. The remaining task is to solve the linear equations involving $M_{\rm stab}$ that enter Eq.~\eqref{eq:number_stab_green}. We form its propagation term by multiplying $V_RV_L$ and rescaling its rows and columns with $D_R$ and $D_L$, respectively. A further pivoted QR factorization gives $M_{\rm stab}=U_MD_MV_M$, so that
\begin{equation}
    M_{\rm stab}^{-1}U_R^\hc
    =V_M^{-1}D_M^{-1}U_M^\hc U_R^\hc.
\end{equation}
To apply this expression, first form the product $U_M^\hc U_R^\hc$, then divide each row by the corresponding entry of $D_M$. The final action of $V_M^{-1}$ is evaluated by solving the triangular equations from the QR factorization and applying the permutation. Multiplication by $U_L^\hc$ completes the reconstruction of $G_{b,\ell}$. These operations evaluate the same Green function as the full cyclic product, with the propagation scales stored separately during accumulation. The second flavor follows from $G_{c,\ell}=\overline{G_{b,\ell}}$.

\section{Nambu Gaussian formalism}
\label{app:paired_gaussian}

Pair creation and annihilation mix the operators $\hat a_i$ and $\hat a_i^+$ during imaginary-time evolution. The Gaussian construction of Sec.~\ref{app:number_conserving_gaussian} therefore extends to a Nambu space containing both sets of operators. We first obtain their propagation matrix, then derive the trace and equal-time correlations, and finally use these results to construct local Monte Carlo updates and their numerical stabilization. The paired implementation and benchmark scripts are available in Ref.~\cite{bafqmc_code}.

\subsection{Nambu algebra}
\label{app:algebra}

For $N$ bosonic modes, collect the annihilation and creation operators into $\hat\chi=(\hat{\bb a},\hat{\bb a}^+)^\trans$. Their canonical commutators take the matrix form
\begin{equation}
    [\hat\chi_\mu,\hat\chi_\nu]=\Omega_{\mu\nu},
    \qquad
    \Omega=
    \begin{pmatrix}
        0&I_N\\
        -I_N&0
    \end{pmatrix}.
\end{equation}
As in the number-conserving calculation, each imaginary-time factor is the exponential of a dimensionless quadratic generator. Including pairing, this generator can be written as
\begin{equation}
    \hat h(A,\xi)=-\frac12\hat\chi^\trans\Omega A\hat\chi+\xi,
    \qquad
    \Omega A+A^\trans\Omega=0.
    \label{eq:hAdef}
\end{equation}
The condition on $A$ makes $\Omega A$ symmetric and ensures that the induced evolution preserves the bosonic commutators. To identify the physical terms, write
\begin{equation}
    A=
    \begin{pmatrix}
        K&\Delta\\
        -\overline{\Delta'}&-K^\trans
    \end{pmatrix},
    \qquad
    \Delta^\trans=\Delta,
    \qquad
    \Delta'^\trans=\Delta'.
\end{equation}
Here $K$ describes number-conserving propagation, while $\Delta$ and $\overline{\Delta'}$ describe pair creation and annihilation. The two pairing matrices are independent for a general auxiliary-field generator. Each is symmetric because the two bosons in a pair obey Bose exchange symmetry. Expanding Eq.~\eqref{eq:hAdef} gives
\begin{equation}
    \hat h(A,\xi)=\hat{\bb a}^+K\hat{\bb a}
    +\frac12\hat{\bb a}^+\Delta(\hat{\bb a}^+)^\trans
    +\frac12\hat{\bb a}^\trans\overline{\Delta'}\hat{\bb a}
    +\xi+\frac12\Tr K.
\label{eq:block_form_h}
\end{equation}
The term $\frac12\Tr K$ arises when the annihilation and creation operators are put in normal order. Thus a quadratic generator with no scalar term in normal order has $\xi=-\frac12\Tr K$. Keeping $\xi$ explicitly retains the scalar contribution to the Monte Carlo weight.

Under quadratic evolution, each creation or annihilation operator becomes a linear combination of these same $2N$ operators. Using $[XY,Z]=X[Y,Z]+[X,Z]Y$ in Eq.~\eqref{eq:hAdef} gives
\begin{equation}
    [\hat h(A,\xi),\hat\chi]=-A\hat\chi.
    \label{eq:appendix_commutator}
\end{equation}
Repeated commutation therefore exponentiates the matrix $A$,
\begin{equation}
    \ee^{\hat h(A,\xi)}\hat\chi\,\ee^{-\hat h(A,\xi)}=\ee^{-A}\hat\chi,
    \qquad
    \hat\chi\,\ee^{\hat h(A,\xi)}=\ee^{\hat h(A,\xi)}\ee^A\hat\chi.
\end{equation}
Passing the operators through successive slices yields
\begin{equation}
    \hat\chi\,\prod_{\ell=1}^{L}\ee^{\hat h(A_\ell,\xi_\ell)}
    =
    \left(\prod_{\ell=1}^{L}\ee^{\hat h(A_\ell,\xi_\ell)}\right)
    \left(\prod_{\ell=1}^{L}\ee^{A_\ell}\right)\hat\chi.
    \label{eq:Mdef}
\end{equation}
Thus the full imaginary-time evolution is encoded by an ordered product of $2N\times2N$ matrices. Each slice satisfies $(\ee^A)^\trans\Omega\ee^A=\Omega$, expressing preservation of the canonical commutators.

\subsection{Trace formula}
\label{app:trace_paired}

For a fixed auxiliary-field configuration, define
\begin{equation}
    \hat U=\prod_{\ell=1}^{L}\ee^{\hat h(A_\ell,\xi_\ell)},
    \qquad
    \mc M=\ee^{A_1}\cdots\ee^{A_L}.
\end{equation}
The trace again sums the weights of all bosonic occupations. To see how this sum is represented in Nambu space, first consider one stable quasiparticle mode $\hat\beta$, including its zero-point contribution, with generator $\hat h=\varepsilon(\hat\beta^+\hat\beta+\frac12)$. The dimensionless exponent $\varepsilon$ is the change in the Boltzmann exponent on adding one quasiparticle, so occupation $n$ contributes $\ee^{(n+1/2)\varepsilon}$. For an oscillator of excitation energy $E>0$, $\varepsilon=-\beta E$. The occupation sum converges for $\operatorname{Re}\varepsilon<0$ and gives
\begin{equation}
    \Tr\,\ee^{\hat h}
    =\sum_{n=0}^{\infty}\ee^{(n+1/2)\varepsilon}
    =\frac{\ee^{\varepsilon/2}}{1-\ee^\varepsilon}.
\end{equation}
In the Nambu basis $(\hat\beta,\hat\beta^+)^\trans$, the same generator is represented by
\begin{equation}
    A=\begin{pmatrix}\varepsilon&0\\0&-\varepsilon\end{pmatrix},
    \qquad \xi=0.
\end{equation}
The annihilation and creation components have reciprocal propagation factors,
\begin{equation}
    \mc M=\begin{pmatrix}\ee^\varepsilon&0\\0&\ee^{-\varepsilon}\end{pmatrix}.
\end{equation}
Thus one physical mode contributes two Nambu eigenvalues. The determinant gives
\begin{equation}
    -\det(I_2-\mc M)
    =\ee^{-\varepsilon}(1-\ee^\varepsilon)^2.
\end{equation}
The square root removes this doubling. Choosing it to reproduce the occupation sum gives
\begin{equation}
    [-\det(I_2-\mc M)]^{1/2}
    =\ee^{-\varepsilon/2}(1-\ee^\varepsilon),
    \qquad
    [-\det(I_2-\mc M)]^{-1/2}
    =\frac{\ee^{\varepsilon/2}}{1-\ee^\varepsilon}.
\end{equation}
This fixes both the magnitude and the sign of the square root from the bosonic trace.

For $N$ independent stable quasiparticle modes, the occupation sums multiply,
\begin{equation}
    \Tr\,\ee^{\sum_j\varepsilon_j(\hat\beta_j^+\hat\beta_j+1/2)}
    =\prod_{j=1}^{N}\frac{\ee^{\varepsilon_j/2}}{1-\ee^{\varepsilon_j}}.
\end{equation}
A canonical change to the quasiparticle basis brings the Nambu propagation matrix to
\begin{equation}
    \mc M=\diag\!\left(
        \ee^{\varepsilon_1},\ldots,\ee^{\varepsilon_N},
        \ee^{-\varepsilon_1},\ldots,\ee^{-\varepsilon_N}
    \right).
\end{equation}
Each physical mode again supplies one reciprocal pair. Multiplying the one-mode identities therefore yields
\begin{equation}
    \left[(-1)^N\det(I_{2N}-\mc M)\right]^{1/2}
    =\exp\!\left(-\frac12\sum_j\varepsilon_j\right)
    \prod_{j=1}^{N}(1-\ee^{\varepsilon_j}).
\end{equation}
Its inverse reproduces the quasiparticle occupation sum, including the zero-point contribution of every mode.

Successive quadratic evolution steps compose into another Gaussian operator, whose Nambu propagation matrix is the ordered product in Eq.~\eqref{eq:Mdef}. The Gaussian trace identity therefore applies to the full evolution, including mixing between modes. For a stable, trace-convergent product, including the scalar factors $\ee^{\xi_\ell}$ from the original generators gives~\cite{munn1978ensemble}
\begin{equation}
    \Tr\,\hat U
    =\exp\!\left(\sum_\ell\xi_\ell\right)
    \left[(-1)^N\det(I_{2N}-\mc M)\right]^{-1/2}.
    \label{eq:paired_trace_formula_appendix}
\end{equation}
The scalar contributions multiply across the slices, and the ordered operator product fixes the square-root sign. Together with the scalar prefactor, the determinant specifies the squared weight. The sign-free criteria complete this information: the TRS proofs in Sec.~\ref{app:trs_proof} express the full weight through reduced determinants, and reflection positivity establishes a non-negative operator trace in Sec.~\ref{app:rp_proof}. The factor $(-1)^N$ is unity for the even number of modes in the sign-free TRS classes considered here.

For configurations whose bare traces diverge, Sec.~\ref{app:regularization} gives the finite-part interpretation of this weight. When every slice separates into the same two independent conjugate bosonic subsystems, their traces can be regularized separately and multiplied to retain a modulus-square weight. This connects the Nambu trace formula to the regularized weights used in the symmetry criteria.

\subsection{Equal-time Nambu Green function}
\label{app:green_paired}

Pairing introduces correlations such as $\langle\hat a_i\hat a_j\rangle_\phi$ alongside the number-conserving correlations $\langle\hat a_i^+\hat a_j\rangle_\phi$. The Nambu Green function collects all of them in one matrix,
\begin{equation}
    G_{\mu\nu}
    =\langle\hat\chi_\mu\hat\chi_\nu\rangle_\phi
    =\frac{\Tr(\hat\chi_\mu\hat\chi_\nu\hat U)}{\Tr\,\hat U}.
    \label{eq:Gdef}
\end{equation}
Its four $N\times N$ blocks have entries
\begin{equation}
    G=
    \begin{pmatrix}
        [\langle\hat a_i\hat a_j\rangle_\phi]_{ij}&[\langle\hat a_i\hat a_j^+\rangle_\phi]_{ij}\\
        [\langle\hat a_i^+\hat a_j\rangle_\phi]_{ij}&[\langle\hat a_i^+\hat a_j^+\rangle_\phi]_{ij}
    \end{pmatrix}.
\end{equation}
The upper-right block contains the correlations $\langle\hat a_i\hat a_j^+\rangle_\phi$ used to define the Green function in Sec.~\ref{app:green_number}, and the lower-left block contains $\langle\hat a_i^+\hat a_j\rangle_\phi$. The diagonal blocks describe pair annihilation and creation. To determine them, use Eq.~\eqref{eq:Mdef} to pass a Nambu operator through the full imaginary-time evolution,
\begin{equation}
    \hat\chi\hat U=\hat U\mc M\hat\chi.
    \label{eq:chiUappendix}
\end{equation}
Multiplying component $\mu$ by $\hat\chi_\nu$ on the right and taking the trace gives
\begin{equation}
    \frac{\Tr(\hat\chi_\mu\hat U\hat\chi_\nu)}{\Tr\hat U}
    =(\mc M G)_{\mu\nu}.
\end{equation}
Cyclicity places the two Nambu operators next to each other. Exchanging their order then introduces the canonical commutator,
\begin{equation}
    \frac{\Tr(\hat U\hat\chi_\nu\hat\chi_\mu)}{\Tr\hat U}
    =G_{\mu\nu}-\Omega_{\mu\nu},
\end{equation}
where we used $[\hat\chi_\mu,\hat\chi_\nu]=\Omega_{\mu\nu}$. Hence $\mc M G=G-\Omega$, or
\begin{equation}
    (I_{2N}-\mc M)G=\Omega,
    \qquad
    G=(I_{2N}-\mc M)^{-1}\Omega.
    \label{eq:Gformula}
\end{equation}
Thus the full set of normal and pair correlations follows from the same propagation matrix that determines the weight. Higher-order observables follow by Wick contraction, including the pair correlations.

\subsection{Local updates and Metropolis ratio}
\label{app:update_paired}

\emph{Weight ratio.} A local auxiliary-field proposal changes both the Nambu propagation and, when present, the scalar term $\xi_\ell$. As in Sec.~\ref{app:update_number}, choose the equal-time cut immediately before the onsite factor being updated, so that its change multiplies the propagation from the left. Write
\begin{equation}
    \ee^{A_\ell'}=(I_{2N}+\Delta_\ell)\ee^{A_\ell},\qquad
    \delta\xi_\ell=\xi_\ell'-\xi_\ell.
    \label{eq:delta_def}
\end{equation}
The propagation beginning at this cut is the cyclic product
\begin{equation}
    \mc M_\ell=\ee^{A_\ell}\ee^{A_{\ell+1}}\cdots\ee^{A_L}\ee^{A_1}\cdots\ee^{A_{\ell-1}}.
\end{equation}
The updated cyclic product is $\mc M_\ell'=(I_{2N}+\Delta_\ell)\mc M_\ell$. Therefore
\begin{equation}
    \frac{W'}{W}
    =
    \ee^{\delta\xi_\ell}
    \left[
    \frac{\det(I_{2N}-\mc M_\ell')}
         {\det(I_{2N}-\mc M_\ell)}
    \right]^{-1/2}.
    \label{eq:update_ratio_paired_start}
\end{equation}
Now factor the matrix in the numerator:
\begin{equation}
\begin{aligned}
    I_{2N}-\mc M_\ell'
    &= I_{2N}-(I_{2N}+\Delta_\ell)\mc M_\ell \\
    &= I_{2N}-\mc M_\ell-\Delta_\ell\mc M_\ell \\
    &=
    \left[
    I_{2N}
    +\Delta_\ell\left(I_{2N}-(I_{2N}-\mc M_\ell)^{-1}\right)
    \right]
    (I_{2N}-\mc M_\ell).
\end{aligned}
    \label{eq:paired_factorization}
\end{equation}
The last line follows from
$[I_{2N}-(I_{2N}-\mc M_\ell)^{-1}](I_{2N}-\mc M_\ell)=-\mc M_\ell$. Substituting Eq.~\eqref{eq:paired_factorization} into Eq.~\eqref{eq:update_ratio_paired_start} gives
\begin{equation}
    \frac{W'}{W}
    =
    \ee^{\delta\xi_\ell}
    \det\left[
    I_{2N}
    +\Delta_\ell\left(I_{2N}-(I_{2N}-\mc M_\ell)^{-1}\right)
    \right]^{-1/2}.
    \label{eq:update_ratio_main}
\end{equation}
The square root inherits the sign fixed for each full weight in Sec.~\ref{app:trace_paired}. In the paired benchmarks, the RP criterion and the trace convergence established in Sec.~\ref{app:benchmark_details} ensure that $W$ and $W'$ are non-negative, fixing the positive ratio for nonzero weights. The identity update, $\Delta_\ell=0$ and $\delta\xi_\ell=0$, gives $W'/W=1$.

\emph{Onsite density updates.} A local density-field proposal acts on the four onsite Nambu components $\hat\chi_i=(\hat b_i,\hat c_i,\hat b_i^+,\hat c_i^+)^\trans$, even when the propagation between updates includes pairing. We retain the total- and relative-density channels of Eqs.~\eqref{eq:benchmark_hs_total} and~\eqref{eq:benchmark_hs_relative}, with $U_2=0$ for the main-text model. Their coefficients are
\begin{equation}
    \lambda_{1,i\ell}=\ii\alpha_1\varphi_{i\ell},\qquad
    \lambda_{2,i\ell}=\alpha_2\phi_{i\ell}.
\end{equation}
The onsite potentials of Sec.~\ref{app:update_number} are therefore $v_{b,i\ell}=\lambda_{2,i\ell}+\lambda_{1,i\ell}$ and $v_{c,i\ell}=\lambda_{2,i\ell}-\lambda_{1,i\ell}$. In the Nambu representation, a density coefficient multiplies $\hat n+\frac12$, as in Eq.~\eqref{eq:block_form_h}. Writing the onsite density terms in this form gives
\begin{equation}
\begin{aligned}
    &\lambda_{2,i\ell}(\hat n_{b,i}+\hat n_{c,i})
    +\lambda_{1,i\ell}(\hat n_{b,i}-\hat n_{c,i}) \\
    &\quad =
    (\lambda_{2,i\ell}+\lambda_{1,i\ell})\left(\hat n_{b,i}+\frac12\right)
    +(\lambda_{2,i\ell}-\lambda_{1,i\ell})\left(\hat n_{c,i}+\frac12\right)
    -\lambda_{2,i\ell}.
\end{aligned}
    \label{eq:nambu_hs_constant}
\end{equation}
The two $\frac12$ terms contribute $\lambda_{2,i\ell}$ in total. The scalar $-\lambda_{2,i\ell}$ cancels this contribution and enters $\xi_\ell$. Thus a local $U_2$-field proposal $\phi_{i\ell}\to\phi_{i\ell}'$ has
\begin{equation}
    R_{\rm const}=\ee^{\delta\xi_\ell}=\exp[-\alpha_2(\phi_{i\ell}'-\phi_{i\ell})],
\end{equation}
whereas a local $U_1$-field proposal $\varphi_{i\ell}\to\varphi_{i\ell}'$ has $R_{\rm const}=1$.

The determinant ratio can also be evaluated using only these four onsite components. For this calculation it is convenient to store
\begin{equation}
    \widetilde G_\ell=(I_{2N}-\mc M_\ell)^{-1},
    \qquad G_\ell=\widetilde G_\ell\Omega.
\end{equation}
The matrix $\widetilde G_\ell$ plays the same role in the local-update formulas as $(I_N-\mc B_\ell)^{-1}$ in Sec.~\ref{app:update_number}. Multiplication by $\Omega$ gives the Nambu correlations. Let $\ket{i,\rho}$ be the basis vector for site $i$ and component $\rho\in\mathcal I=\{b,c,b^+,c^+\}$ in the $2N$-dimensional Nambu space. Collecting these four basis vectors gives
\begin{equation}
    S_i=(\ket{i,b},\ket{i,c},\ket{i,b^+},\ket{i,c^+}).
\end{equation}
The columns of $S_i$ insert the onsite components into the full system, and $S_i^\hc$ extracts them. Since the proposal changes only these components, its update matrix is
\begin{equation}
    \Delta_\ell=S_iD_iS_i^\hc
    =\sum_{\rho,\sigma\in\mathcal I}\ket{i,\rho}(D_i)_{\rho\sigma}\bra{i,\sigma},
\end{equation}
where $D_i$ is a $4\times4$ matrix in the onsite Nambu basis. The determinant matrix in Eq.~\eqref{eq:update_ratio_main} is
\begin{equation}
    C_\ell=I_{2N}+\Delta_\ell(I_{2N}-\widetilde G_\ell)
    =
    I_{2N}+S_iD_i S_i^\hc(I_{2N}-\widetilde G_\ell).
\end{equation}
The matrix determinant lemma then reduces the full determinant to a $4\times4$ determinant,
\begin{equation}
    \det C_\ell
    =
    \det\!\left[
    I_4+D_i S_i^\hc(I_{2N}-\widetilde G_\ell)S_i
    \right]
    \equiv \det P_i.
    \label{eq:paired_local_det_lemma}
\end{equation}
In components,
\begin{equation}
    (P_i)_{\rho\sigma}
    =
    \delta_{\rho\sigma}
    +\sum_{\tau\in\mathcal I}
    (D_i)_{\rho\tau}
    \left[
    \delta_{\tau\sigma}
    -\bra{i,\tau}\widetilde G_\ell\ket{i,\sigma}
    \right],
    \qquad
    \rho,\sigma\in\mathcal I.
    \label{eq:paired_local_block}
\end{equation}
A density field rescales each onsite operator, with inverse factors for creation and annihilation. Thus $D_i$ is diagonal in this basis. For a local $U_1$ update $\varphi_{i\ell}\to\varphi_{i\ell}'$ with $\delta\lambda_1=\ii\alpha_1(\varphi_{i\ell}'-\varphi_{i\ell})$,
\begin{equation}
    D_i^{(1)}=
    \begin{pmatrix}
        \ee^{\delta\lambda_1}-1&0&0&0\\
        0&\ee^{-\delta\lambda_1}-1&0&0\\
        0&0&\ee^{-\delta\lambda_1}-1&0\\
        0&0&0&\ee^{\delta\lambda_1}-1
    \end{pmatrix},
\end{equation}
while for a local $U_2$ update $\phi_{i\ell}\to\phi_{i\ell}'$ with $\delta\lambda_2=\alpha_2(\phi_{i\ell}'-\phi_{i\ell})$,
\begin{equation}
    D_i^{(2)}=
    \begin{pmatrix}
        \ee^{\delta\lambda_2}-1&0&0&0\\
        0&\ee^{\delta\lambda_2}-1&0&0\\
        0&0&\ee^{-\delta\lambda_2}-1&0\\
        0&0&0&\ee^{-\delta\lambda_2}-1
    \end{pmatrix}.
\end{equation}
The determinant contribution to a local proposal is $(\det P_i)^{-1/2}$, with the positive square root for the RP ensemble specified above. Combining it with the Gaussian HS measure and the normal-ordering constant gives the full local Metropolis ratio
\begin{equation}
    R=
    \exp\left[-\frac12(x_{i\ell}'^{\,2}-x_{i\ell}^{2})\right]
    R_{\rm const}
    (\det P_i)^{-1/2},
    \qquad
    p(x_{i\ell}\to x_{i\ell}')=\min(1,R),
    \label{eq:paired_metropolis_ratio}
\end{equation}
where $x_{i\ell}$ denotes the real field being updated, either $\phi_{i\ell}$ or $\varphi_{i\ell}$, and the proposal distribution is symmetric as in Sec.~\ref{app:update_number}. For the main-text model, only the relative-density $U_1$ field is sampled and $R_{\rm const}=1$.

\emph{Green-function update.} Once the proposal is accepted, the correlations throughout the system must be updated. The same onsite matrix $P_i$ determines this change. From Eq.~\eqref{eq:paired_factorization},
\begin{equation}
    \widetilde G_\ell'
    =
    \widetilde G_\ell C_\ell^{-1}.
\end{equation}
Applying the Woodbury identity to $C_\ell=I_{2N}+S_iD_i S_i^\hc(I_{2N}-\widetilde G_\ell)$ gives
\begin{equation}
    C_\ell^{-1}
    =
    I_{2N}
    -
    S_i P_i^{-1}D_i S_i^\hc(I_{2N}-\widetilde G_\ell).
    \label{eq:paired_woodbury}
\end{equation}
Multiplying by $\widetilde G_\ell$ gives the rank-four correction
\begin{equation}
    \widetilde G_\ell'
    =\widetilde G_\ell
    -\widetilde G_\ell S_i P_i^{-1}D_i S_i^\hc(I_{2N}-\widetilde G_\ell).
\end{equation}
The columns of $\widetilde G_\ell S_i$ connect the updated site to the rest of the system, while $S_i^\hc(I_{2N}-\widetilde G_\ell)$ supplies the corresponding rows. Their product spreads the local change to all matrix elements. In components,
\begin{equation}
    \widetilde G_\ell'
    =
    \widetilde G_\ell
    -
    \sum_{\rho,\sigma,\tau\in\mathcal I}
    \widetilde G_\ell\ket{i,\rho}
    (P_i^{-1})_{\rho\sigma}(D_i)_{\sigma\tau}
    \left(\bra{i,\tau}-\bra{i,\tau}\widetilde G_\ell\right).
    \label{eq:paired_green_update}
\end{equation}
Each proposal therefore requires a four-dimensional determinant for the acceptance probability. On acceptance, Eq.~\eqref{eq:paired_green_update} updates the full resolvent through the same local block, and $G_\ell'=\widetilde G_\ell'\Omega$ gives the updated physical correlations.

\subsection{Numerical stabilization}
\label{app:stabilization_paired}

The local updates require an accurate equal-time resolvent $\widetilde G_\ell=(I_{2N}-\mc M_\ell)^{-1}$. Pairing mixes creation and annihilation components, whose propagation can develop widely separated scales over long imaginary times. As in Sec.~\ref{app:stabilization_number}, these scales are stored separately during matrix multiplication, and the resolvent is periodically reconstructed from the full propagation.

The cyclic propagation matrix is
\begin{equation}
    \mc M_\ell=\ee^{A_\ell}\ee^{A_{\ell+1}}\cdots\ee^{A_L}\ee^{A_1}\cdots\ee^{A_{\ell-1}}.
\end{equation}
For two flavors on $N_s$ sites, $N=2N_s$ and these matrices have dimension $4N_s$. Accepted field proposals update $\widetilde G_\ell$ through Eq.~\eqref{eq:paired_green_update}. Moving the equal-time cut to the next slice moves $\ee^{A_\ell}$ from the beginning to the end of the product, giving
\begin{equation}
    \mc M_{\ell+1}=\ee^{-A_\ell}\mc M_\ell\ee^{A_\ell},
    \qquad
    \widetilde G_{\ell+1}=\ee^{-A_\ell}\widetilde G_\ell\ee^{A_\ell}.
\end{equation}
These local updates and propagations accumulate roundoff error. We therefore reconstruct the resolvent after each window of $N_{\rm wrap}$ slices, with $N_{\rm wrap}=10$ in the reported paired benchmarks.

For a window beginning at slice $j$, form
\begin{equation}
    X_j=\ee^{A_j}\ee^{A_{j+1}}\cdots\ee^{A_{j+N_{\rm wrap}-1}},
\end{equation}
with indices understood modulo $L$. The column-pivoted QR factorization $X_jP_j=Q_jR_j$ separates the unitary mixing $Q_j$ from the triangular factor $R_j$. Extracting $(D_j)_{ii}=|(R_j)_{ii}|$ gives
\begin{equation}
    X_j=U_jD_jV_j,
    \qquad U_j=Q_j,
    \qquad V_j=D_j^{-1}R_jP_j^\trans.
\end{equation}
The positive diagonal matrix $D_j$ stores the propagation scales, while $V_j$ contains the normalized triangular factor and the column permutation. Longer intervals are accumulated by the repeated factorization described in Sec.~\ref{app:stabilization_number}: each new window is multiplied into the stored unitary factor, its columns are rescaled by the stored $D$, and a new QR factorization extracts the updated scales.

At a window boundary, reconstruct the two intervals that meet at the equal-time cut,
\begin{equation}
    \ee^{A_\ell}\cdots\ee^{A_L}=U_RD_RV_R,
    \qquad
    \ee^{A_1}\cdots\ee^{A_{\ell-1}}=V_LD_LU_L.
\end{equation}
The second factorization is obtained through the Hermitian-conjugate product, as in Sec.~\ref{app:stabilization_number}. This places the unitary factors at the outer ends of the full cyclic product,
\begin{equation}
    \mc M_\ell=U_RD_RV_RV_LD_LU_L.
\end{equation}
Removing these outer factors from $I_{2N}-\mc M_\ell$ leaves the matrix needed for the Green-function solve,
\begin{equation}
    I_{2N}-\mc M_\ell=U_R M_{\rm stab}U_L,
    \qquad
    M_{\rm stab}=U_R^\hc U_L^\hc-D_RV_RV_LD_L.
\end{equation}
Here $M_{\rm stab}$ is the $2N$-dimensional counterpart of Eq.~\eqref{eq:number_mstab_def}. The reconstructed resolvent is
\begin{equation}
    \widetilde G_\ell=U_L^\hc M_{\rm stab}^{-1}U_R^\hc.
\end{equation}
To evaluate it, first form $V_RV_L$ and rescale its rows and columns by $D_R$ and $D_L$. A further scaled QR factorization, $M_{\rm stab}=U_MD_MV_M$, gives
\begin{equation}
    M_{\rm stab}^{-1}U_R^\hc
    =V_M^{-1}D_M^{-1}U_M^\hc U_R^\hc.
\end{equation}
The solve starts from $U_M^\hc U_R^\hc$, divides each row by the corresponding entry of $D_M$, and applies $V_M^{-1}$ through triangular equations and the stored permutation. Multiplication by $U_L^\hc$ completes the reconstruction. This periodically replaces the locally propagated resolvent with one evaluated from the full factorized product. The same reconstructed matrix supplies both the local acceptance ratios and the physical correlations $G_\ell=\widetilde G_\ell\Omega$.

\section{Gaussian-cutoff regularization of determinant weights}
\label{app:regularization}

Consider a stable many-body Hamiltonian with a finite grand-canonical partition function, $Z=\Tr\,\ee^{-\beta(\hat H-\mu\hat N)}<\infty$. The HS transformation expresses its Boltzmann factor as an integral over quadratic propagators through exact operator identities on each imaginary-time slice,
\begin{equation}
    \ee^{-\beta(\hat H-\mu\hat N)}
    =\lim_{\Delta\tau\to0}
    \int\mc D\phi\;P_{\rm HS}(\phi)\,
    \hat U(\phi),
    \qquad
    \hat U(\phi)=\prod_{\ell=1}^{L}\ee^{\hat h_\ell(\phi_\ell)}.
\end{equation}
The limit removes the Trotter error, while the HS transformation introduces no approximation. The complete auxiliary-field integral therefore recovers the original Boltzmann factor and its finite trace. Nevertheless, for some fixed auxiliary-field configurations, the bare weight $W_{\rm bare}(\phi)=\Tr\hat U(\phi)$ diverges. How do these divergent configuration weights combine under auxiliary-field integration to recover the finite many-body partition function? A Gaussian-cutoff regularization provides an interpretation by separating the finite contribution of each configuration from the divergent terms that cancel upon integration.

For number-conserving evolution, let $r_1(\phi),\ldots,r_N(\phi)$ be the eigenvalues of the single-particle propagation matrix $\mc B(\phi)$. In a diagonal normal-mode basis, occupying mode $i$ with $n_i$ bosons contributes $r_i(\phi)^{n_i}$ to the many-body weight. Summing over all occupations gives the bare trace
\begin{equation}
    W_{\rm bare}(\phi)=\sum_{\bb n\ge0} a_{\bb n}(\phi),
    \qquad
    a_{\bb n}(\phi)=\prod_{i=1}^{N}r_i(\phi)^{n_i},
    \label{eq:bare_occupation_series}
\end{equation}
where $\bb n=(n_1,\ldots,n_N)$ labels the boson occupations. When $|r_i|<1$ for every mode, each occupation sum is a convergent geometric series and Eq.~\eqref{eq:bare_occupation_series} equals $\prod_i(1-r_i)^{-1}$. If $|r_i|>1$ for any mode, its contribution grows with $n_i$ and the bare trace diverges. To identify its finite part, introduce the Gaussian regulator
\begin{equation}
    W_{\Lambda}(\phi)
    =
    \sum_{\bb n\ge0}
    \prod_{i=1}^{N}
    r_i(\phi)^{n_i}
    \exp\!\left[-\frac{|r_i(\phi)|^{2n_i}}{\Lambda^2}\right].
    \label{eq:gaussian_cutoff_weight}
\end{equation}
For finite $\Lambda$, the Gaussian factor renders the growing occupation sums convergent.

The finite part can be identified first for a single mode with eigenvalue $r$ and $|r|>1$. Its regulated occupation sum is
\begin{equation}
    W_{\Lambda}(r)
    =
    \sum_{n\ge0}r^n
    \exp\!\left(-\frac{|r|^{2n}}{\Lambda^2}\right).
\end{equation}
The regulator becomes effective near $|r|^n\sim\Lambda$, corresponding to the occupation scale $n_*(\Lambda)=\ln\Lambda/\ln|r|$. As $\Lambda$ increases, the regulated sum separates into a cutoff-dependent contribution and a finite term,
\begin{equation}
    W_{\Lambda}(r)=P_{\Lambda}(r)+\frac{1}{1-r}+o(1),
    \qquad r\ne1,
\end{equation}
where $P_{\Lambda}$ collects the terms that grow with the cutoff. For $|r|<1$, the regulator can be removed directly, and the finite term $(1-r)^{-1}$ equals the convergent occupation sum. For multiple modes, this separation gives
\begin{equation}
    W_{\Lambda}(\phi)=P_{\Lambda}(\phi)+W_{\rm reg}(\phi)+o(1),
\end{equation}
with
\begin{equation}
    W_{\rm reg}(\phi)
    =
    \prod_{i=1}^{N}\frac{1}{1-r_i(\phi)}
    =
    \det[I_N-\mc B(\phi)]^{-1}.
    \label{eq:reg_weight_appendix}
\end{equation}
The inverse determinant thus represents the cutoff-independent finite part of the bosonic occupation sum.

To relate this finite part to the interacting partition function, retain the regulator until the auxiliary fields have been integrated. Denoting the Fock-space regulator by $R_\Lambda$, the regulated partition function is
\begin{equation}
    Z_{\Lambda}
    =
    \Tr\left[
    R_\Lambda\prod_{\ell}\ee^{-\Delta\tau \hat H}
    \right]
    =
    \int\mc D\phi\;P_{\rm HS}(\phi)\,
    \Tr\left[
    R_\Lambda\prod_{\ell}\ee^{\hat h_\ell(\phi_\ell)}
    \right].
\end{equation}
As $\Lambda\to\infty$, the left-hand side recovers the finite physical partition function. In the auxiliary-field representation, each regulated trace contains the cutoff-dependent contribution $P_\Lambda(\phi)$ and the finite part $W_{\rm reg}(\phi)$. Since the physical partition function is independent of the cutoff in this limit, the cutoff-dependent divergent terms must cancel exactly upon integration over the complete auxiliary-field domain. Only the finite, regularized contributions remain, giving
\begin{equation}
    Z=\int\mc D\phi\;P_{\rm HS}(\phi)\,W_{\rm reg}(\phi).
\end{equation}

For evolution with pairing, the quasiparticle occupation sum has the same structure. In the quasiparticle basis of Sec.~\ref{app:trace_paired}, $\ee^{\varepsilon_j(\phi)}$ is the factor associated with adding one quasiparticle to mode $j$, and
\begin{equation}
    W_{\rm bare}(\phi)
    =
    \exp\!\left[\sum_{\ell=1}^{L}\xi_\ell(\phi_\ell)+\frac12\sum_j\varepsilon_j(\phi)\right]
    \sum_{n_1,\ldots,n_N\ge0}
    \prod_{j=1}^{N}\ee^{n_j\varepsilon_j(\phi)}.
\end{equation}
The factor $\exp(\frac12\sum_j\varepsilon_j)$ accounts for the zero-point contribution of the quasiparticle modes, while $\exp(\sum_\ell\xi_\ell)$ retains the scalar terms in the original generators. Regularizing the quasiparticle occupation sums gives the finite part
\begin{equation}
    W_{\rm reg}(\phi)
    =
    \exp\!\left[\sum_{\ell=1}^{L}\xi_\ell(\phi_\ell)+\frac12\sum_j\varepsilon_j(\phi)\right]
    \prod_{j=1}^{N}\frac{1}{1-\ee^{\varepsilon_j(\phi)}}.
\end{equation}
In Nambu notation this finite part is written as
\begin{equation}
    W_{\rm reg}(\phi)=\exp\!\left[\sum_{\ell=1}^{L}\xi_\ell(\phi_\ell)\right]\left[(-1)^N\det[I_{2N}-\mc M(\phi)]\right]^{-1/2},
\end{equation}
where the scalar prefactor and the choice of square root are fixed as in Sec.~\ref{app:trace_paired}. For trace-convergent RP products, reflection positivity fixes the non-negative operator weight.

The determinant expressions in Eqs.~\eqref{eq:trace_formula_number} and~\eqref{eq:trace_formula_main} therefore give the regularized weights $W_{\rm reg}(\phi)$ sampled in BAFQMC. Wherever the bare trace converges, $W_{\rm reg}(\phi)=W_{\rm bare}(\phi)$. For configurations with divergent bare traces, the determinant expressions represent the finite parts identified above.

When time reversal separates every slice into two independent conjugate subsystems, as in the pairing class proved in Sec.~\ref{app:trs_pairing}, the cutoff is applied to the two subsystems separately. The Fock space and slice generators decompose as $\mathcal F=\mathcal F_+\otimes\mathcal F_-$ and $\hat h_\ell=\hat h_{\ell,+}+\hat h_{\ell,-}$, with half of the real scalar $\xi_\ell$ assigned to each subsystem. Using identical cutoffs for each state and its time-reversed partner preserves the conjugacy of the regulated traces,
\begin{equation}
    W_{\Lambda,\pm}=\Tr_{\mathcal F_\pm}
    \left[R_{\Lambda,\pm}\prod_\ell\ee^{\hat h_{\ell,\pm}}\right],
    \qquad W_{\Lambda,-}=\overline{W_{\Lambda,+}}.
\end{equation}
Subtracting the cutoff-dependent growing terms separately from each trace preserves $W_-=\overline{W_+}$ for their finite parts. Multiplying these finite parts gives
\begin{equation}
    W_{\rm reg}=W_+\overline{W_+}=|W_+|^2
    =\ee^{\sum_\ell\xi_\ell}
    \left|\det(I_N-\mc M_+)^{-1}\right|,
    \label{eq:reg_conjugate_sectors}
\end{equation}
where $\mc M_+$ is the Nambu propagation matrix of one subsystem, containing $N/2$ bosonic modes. The last equality follows by taking the absolute value of its squared Gaussian trace. Thus conjugate subsystems retain a non-negative weight even when their bare traces diverge. This prescription agrees with the ordinary trace wherever it converges. When the evolution cannot be separated into two independent conjugate subsystems, this separate regularization of conjugate traces does not apply.

The TRS criteria proved in Sec.~\ref{app:trs_proof} apply to these regularized weights and do not require the bare trace to converge for every fixed auxiliary-field configuration. The Kramers and split-orthogonal proofs fix the sign directly through their reduced determinants.

\section{Proof of the time-reversal criterion}
\label{app:non-negative_proofs}
\label{app:trs_proof}

Time-reversal symmetry (TRS) ensures the positivity of Boltzmann weights that might otherwise acquire an unphysical sign or complex phase. In this section, we provide detailed proofs demonstrating how the identified symmetry classes rigorously eliminate the sign problem. We establish positivity for the three fundamental classes in the order introduced in the main text, beginning with the familiar Kramers argument for number-conserving particles~\cite{wu2005prb,li2016prl}.

Following the formalism of Sec.~\ref{app:algebra}, we express the Hamiltonian slices and the propagation matrix using Nambu operators:
\begin{equation}
    \hat h_\ell=-\frac12\hat\chi^\trans\Omega A_\ell\hat\chi+\xi_\ell,
    \qquad \mc M=\ee^{A_1}\cdots\ee^{A_L}.
    \label{eq:trs_configuration_weight}
\end{equation}
The antiunitary time reversal $\Theta$ complex-conjugates numerical coefficients, preserves bosonic commutators, and maps $\hat d^+$ to $(\Theta\hat d)^+$. Representing its action on the Nambu basis as $\Theta(\hat\chi)=U\hat\chi$, these requirements—combined with the condition of slice invariance—impose the following constraints:
\begin{equation}
    U^\trans\Omega U=\Omega,\qquad
    C\overline U C=U,\qquad
    U^{-1}\overline{A_\ell}U=A_\ell,\qquad \xi_\ell\in\mathbb R,
    \quad C=\begin{pmatrix}0&I_N\\I_N&0\end{pmatrix}.
    \label{eq:trs_physical_TR_conditions}
\end{equation}
Here, the matrix $C$ exchanges creation and annihilation operators. By definition, all TRSs within a given class leave every time slice invariant and mutually anticommute.

\subsection{Kramers class \texorpdfstring{\(\{-,-\}\)}{\{-,-\}}}
\label{app:trs_kramers}
\label{app:trs_number}

For number-conserving slices $\hat h_\ell=\hat{\bb a}^+B_\ell\hat{\bb a}$, suppose $\Theta(\hat{\bb a})=U\hat{\bb a}$ with unitary $U$ and $\Theta^2=-I$. Every slice and their ordered product obey
\begin{equation}
    U^\hc\overline{B_\ell}U=B_\ell,\qquad
    U^\hc\overline{\mc B}U=\mc B,\qquad
    \mc B=\ee^{B_1}\cdots\ee^{B_L}.
    \label{eq:Bsym_prod}
\end{equation}
If $\mc Bv=\lambda v$, its time-reversed partner satisfies
\begin{equation}
    \mc B(\overline U\,\overline v)=\overline\lambda\,(\overline U\,\overline v).
\end{equation}
For real $\lambda$, the partners remain independent. If the partner were $cv$, applying time reversal twice would give $-v=|c|^2v$. Thus complex eigenvalues occur in conjugate pairs and real eigenvalues have Kramers degeneracy. Each pair contributes a non-negative factor to the determinant,
\begin{equation}
    \det(I_N-\mc B)=\prod_{\rm pairs}|1-\lambda|^2\ge0,
    \qquad W=\det(I_N-\mc B)^{-1}>0.
    \label{eq:trs_det_positive_sm}
\end{equation}
The same conclusion holds at eigenvalue coalescences by continuity of the determinant. This is the bosonic version of the time-reversal argument in Ref.~\cite{wu2005prb}.

Particle-number conservation supplies a second time reversal in Nambu space. Composing $\Theta_1=\Theta$ with $\hat{\bb a}\mapsto\ii\hat{\bb a}$ gives $\Theta_2^2=-I$ and $\Theta_1\Theta_2=-\Theta_2\Theta_1$. The Kramers class retains this pair even when physical particle number is not conserved. Define
\begin{equation}
    Q=\Theta_1\Theta_2,\qquad Q^2=-I,\qquad [Q,A_\ell]=0.
    \label{eq:trs_Q_properties}
\end{equation}
The corresponding quadratic charge is conserved,
\begin{equation}
    \hat Q=-\frac12\hat\chi^\trans\Omega(\ii Q)\hat\chi,
    \qquad [\hat Q,\hat h_\ell]=0.
    \label{eq:trs_charge_conservation}
\end{equation}
Its two operator sectors have $Q=\pm\ii$. An operator $\hat d$ and its adjoint have opposite eigenvalues, since $Q\hat d=\ii\hat d$ implies $Q\hat d^+=-\ii\hat d^+$. We can therefore describe propagation using the $N$ operators in the $Q=+\ii$ sector, with the other sector fixed by Hermitian conjugation. Write their slice and full propagation matrices as $A_{\ell,+\ii}$ and $\mc M_{+\ii}$.

Time reversal acts within the selected sector. Its complex conjugation cancels the sign from anticommuting with $Q$,
\begin{equation}
    Q(\Theta_1\hat d)=-\Theta_1(Q\hat d)
    =-\Theta_1(\ii\hat d)=\ii\Theta_1\hat d.
    \label{eq:trs_Kramers_in_reduced_space}
\end{equation}
The same Kramers argument therefore gives
\begin{equation}
    \det(I_N-\mc M_{+\ii})=\prod_{\rm pairs}|1-\lambda|^2\ge0,
    \qquad \Tr A_{\ell,+\ii}\in\mathbb R.
    \label{eq:trs_Kramers_determinant}
\end{equation}

One bosonic sign remains to be accounted for. The selected operators may include creation operators as well as annihilation operators. Normalize their commutators as
\begin{equation}
    [\hat d_\mu,\hat d_\nu^+]=D_{\mu\nu},\qquad
    D=\diag(I_{N-N_-},-I_{N_-}),
    \label{eq:trs_reduced_commutator}
\end{equation}
where $N_-$ counts creation directions. The sign is easiest to understand for one diagonal bosonic mode. Let $\ee^x$ be the propagation factor for adding one boson to this mode over the full imaginary-time evolution. Including the zero-point contribution, the mode weight is
\begin{equation}
    \sum_{n=0}^{\infty}\ee^{(n+1/2)x}
    =\frac{\ee^{x/2}}{1-\ee^x}.
\end{equation}
In Nambu language the same oscillator appears as a reciprocal pair of directions, with propagation factors $\ee^x$ and $\ee^{-x}$. Choosing the creation direction reverses $x$ and gives
\begin{equation}
    \frac{\ee^{-x/2}}{1-\ee^{-x}}
    =-\frac{\ee^{x/2}}{1-\ee^x}.
    \label{eq:trs_boson_reversed_mode}
\end{equation}
Thus the same physical oscillator acquires one minus sign when its creation direction is used in the reduced sector. Each selected creation direction therefore contributes one minus sign. The same relation extends to the regularized determinant representation. For the full product, preservation of the canonical commutators fixes the complementary Nambu block to $D(\mc M_{+\ii}^\trans)^{-1}D$, giving
\begin{equation}
    (-1)^N\det(I_{2N}-\mc M)
    =\det(\mc M_{+\ii})^{-1}\det(I_N-\mc M_{+\ii})^2.
    \label{eq:trs_boson_reduced_square}
\end{equation}
Together with $\det\mc M_{+\ii}=\exp(\sum_\ell\Tr A_{\ell,+\ii})$ and the oscillator signs above, the Gaussian trace formula yields
\begin{equation}
    W=(-1)^{N_-}
    \exp\!\left[\sum_\ell\left(\xi_\ell+\frac12\Tr A_{\ell,+\ii}\right)\right]
    \det(I_N-\mc M_{+\ii})^{-1}.
    \label{eq:trs_boson_reduced_weight}
\end{equation}
This reduction applies to any conserved $Q$ with $Q^2=-I$ and will also be used for the split-orthogonal class.

For the Kramers class, every creation direction has an independent time-reversed partner with the same commutator sign,
\begin{equation}
    [\Theta_1\hat d,(\Theta_1\hat d)^+]=[\hat d,\hat d^+],
    \qquad [\hat d,(\Theta_1\hat d)^+]=0.
    \label{eq:trs_commutator_Kramers_pair}
\end{equation}
The second equality follows by applying $\Theta_1$ once more and using $\Theta_1^2=-I$. Consequently $N_-$ is even. The exponential in Eq.~\eqref{eq:trs_boson_reduced_weight} is positive and the determinant is non-negative, establishing $W>0$. The conserved charge thus allows the Kramers pairing argument to apply to general quadratic evolution.

\subsection{Pairing class \texorpdfstring{\(\{+,-\}\)}{\{+,-\}}}
\label{app:trs_pairing}

Assume that every time slice preserves two time-reversal symmetries subject to the constraints
\begin{equation}
    \Theta_1^2=I,\qquad \Theta_2^2=-I,\qquad
    \Theta_1\Theta_2=-\Theta_2\Theta_1.
\end{equation}
Because $\Theta_1$ and $\Theta_2$ are both antiunitary, their product constitutes a unitary involution,
\begin{equation}
    P=\Theta_1\Theta_2,\qquad P^2=I,\qquad
    [P,A_\ell]=0,\qquad P\Theta_1=-\Theta_1P.
\end{equation}
The commutation relation $[P,A_\ell]=0$ ensures that the $P=+1$ and $P=-1$ eigenspaces decouple and evolve independently. Since $P$ commutes with Hermitian conjugation, any operator and its adjoint must reside in the same sector. Moreover, operators belonging to opposite sectors strictly commute. For instance, given $P\hat d_+=\hat d_+$ and $P\hat d_-=-\hat d_-$, preserving the canonical bosonic algebra gives
\begin{equation}
    [\hat d_+,\hat d_-]=[P\hat d_+,P\hat d_-]
    =-[\hat d_+,\hat d_-]=0.
\end{equation}
An operator can thus only have a canonical conjugate within its own sector, implying that each sector describes a complete, independent bosonic subsystem. The total Fock space and the Hamiltonian therefore decompose as
\begin{equation}
    \mathcal F=\mathcal F_+\otimes\mathcal F_-,\qquad
    \hat h_\ell=\hat h_{\ell,+}+\hat h_{\ell,-}.
\end{equation}
Because $P$ and $\Theta_1$ anticommute, the time-reversal operation $\Theta_1$ exchanges these two subsystems. Denoting their respective statistical weights by $W_+$ and $W_-$, the application of $\Theta_1$ ensures that $W_-=\overline{W_+}$. The total weight is therefore inherently non-negative:
\begin{equation}
    W=W_+W_-=|W_+|^2\ge0.
    \label{eq:trs_pairing_weight}
\end{equation}
In cases involving divergent traces, the paired cutoffs introduced in Sec.~\ref{app:regularization} preserve this conjugacy relation between their finite parts. Consequently, Eq.~\eqref{eq:reg_conjugate_sectors} yields a strictly positive determinantal weight. This pairing class is therefore the bosonic counterpart to the fermionic Majorana class identified in Ref.~\cite{li2016prl}. Pairing interactions within either subsystem remain fully compatible with this conjugacy relation, preserving the positivity of the total weight.

\subsection{Split-orthogonal class \texorpdfstring{\(\{+,+,+\}\)}{\{+,+,+\}}}

Consider three mutually anticommuting time-reversal symmetries satisfying $\Theta_1^2=\Theta_2^2=\Theta_3^2=I$. Their product $Q=\Theta_1\Theta_2$ obeys $Q^2=-I$ and commutes with every time slice. As before, we project onto the selected $Q=+\ii$ sector. The proof has two ingredients: the third symmetry balances creation and annihilation directions in this sector, while the resulting propagation matrix belongs to a split-orthogonal group whose determinant sign is fixed.

Let us first choose a basis of canonical Nambu operators invariant under $\Theta_1$, which can be constructed using combinations of the form $\hat d+\Theta_1\hat d$ and $\ii(\hat d-\Theta_1\hat d)$. In this basis, the symmetries take the form
\begin{equation}
    \Theta_1=\mc K,\qquad \Theta_2=\ii R_2\mc K,\qquad
    \Theta_3=\ii R_3\mc K,
    \label{eq:trs_three_plus_basis}
\end{equation}
where $R_2$ and $R_3$ are real matrices and $\mc K$ denotes complex conjugation. The factors of $\ii$ arise necessarily from the anticommutation relations with $\Theta_1$. Invariance of each time slice and preservation of the bosonic algebra impose the following constraints:
\begin{gather}
    \overline{A_\ell}=A_\ell,\qquad [R_2,A_\ell]=[R_3,A_\ell]=0,
    \qquad R_2R_3=-R_3R_2,\notag\\
    R_j^2=I,\qquad R_j^\trans\Omega R_j=-\Omega,\qquad
    CR_jC=-R_j,\qquad j=2,3.
    \label{eq:trs_three_plus_physical}
\end{gather}
Because $Q=-\ii R_2$, the two eigenspaces of $Q$ admit purely real bases. Consequently, both the generator $A_{\ell,+\ii}$ and the propagation matrix $\mc M_{+\ii}$ in the selected sector are real-valued.

As in the Kramers class, we describe propagation using the $N$ operators in the $Q=+\ii$ sector, with slice and full propagation matrices $A_{\ell,+\ii}$ and $\mc M_{+\ii}$. If $\hat d$ is an annihilation direction in the selected $Q=+\ii$ sector, then $\Theta_3(\hat d^+)$ is a creation direction in the same sector. Their commutators have opposite signs:
\begin{equation}
    [\Theta_3(\hat d^+),(\Theta_3(\hat d^+))^+]=-[\hat d,\hat d^+].
\end{equation}
Because $\Theta_3$ is invertible, this pairing is one-to-one. The selected $Q=+\ii$ sector therefore contains $N/2$ annihilation and $N/2$ creation directions, so $N_-=N/2$. Substituting this into Eq.~\eqref{eq:trs_boson_reduced_weight} gives the complete Boltzmann weight
\begin{equation}
    W=(-1)^{N/2}
    \exp\!\left[\sum_\ell\left(\xi_\ell+\frac12\Tr A_{\ell,+\ii}\right)\right]
    \det(I_N-\mc M_{+\ii})^{-1}.
    \label{eq:trs_boson_three_plus_reduced_weight}
\end{equation}
The scalar exponential is positive. It remains to determine the sign of $(-1)^{N/2}\det(I_N-\mc M_{+\ii})^{-1}$.

We determine this sign by identifying the structure preserved by $\mc M_{+\ii}$. Define the real symmetric matrix $\eta$ on the selected $Q=+\ii$ sector by
\begin{equation}
    \eta=(\Omega R_3)\big|_{Q=+\ii}.
    \label{eq:trs_eta_definition}
\end{equation}
Here, the restriction denotes evaluating the bilinear form within the $Q=+\ii$ sector. Equation~\eqref{eq:trs_three_plus_physical} implies $(\Omega R_3)^\trans=\Omega R_3$. Because $R_3$ commutes with every time slice, the generator $A_{\ell,+\ii}$ and the propagation matrix $\mc M_{+\ii}$ preserve this metric:
\begin{equation}
    A_{\ell,+\ii}^\trans\eta+\eta A_{\ell,+\ii}=0,\qquad
    \mc M_{+\ii}^\trans\eta\mc M_{+\ii}=\eta.
    \label{eq:trs_eta_propagation}
\end{equation}
The operator $CR_3$ preserves the $Q=+\ii$ sector and anticommutes with $\eta$, transforming $\eta$ into $-\eta$. Thus every positive-$\eta$ eigenvector is paired with a negative-$\eta$ eigenvector, and vice versa. Since $CR_3$ is invertible, these pairings are one-to-one, so $\eta$ has equal numbers of positive and negative eigenvalues. A real change of basis then brings $\eta$ to the canonical form
\begin{equation}
    \eta=\diag(I_{N/2},-I_{N/2}).
\end{equation}
In this basis, Eq.~\eqref{eq:trs_eta_propagation} identifies $\mc M_{+\ii}$ as an element of the split-orthogonal group $O(N/2,N/2)$.

The split-orthogonal determinant theorem~\cite{wang2015prl} states that, if $\mc M$ belongs to the split-orthogonal group $O(N/2,N/2)$,
\begin{equation}
    \det(I_N+\mc M)
    \begin{cases}
        \ge 0, & \text{if }\mc M\in O^{++}(N/2,N/2),\\
        \le 0, & \text{if }\mc M\in O^{--}(N/2,N/2),\\
        =0, & \text{otherwise}.
    \end{cases}
\end{equation}
Here $O^{\pm\pm}(N/2,N/2)$ denotes the component labeled by the signs of the determinants of the two diagonal $N/2\times N/2$ blocks of $\mc M$. Because the physical time evolution is continuously connected to the identity (via $\ee^{sA_{\ell,+\ii}}$ for $0\le s\le1$), the two $N/2 \times N/2$ diagonal blocks of $\mc M_{+\ii}$ have strictly positive determinants. However, our formula requires the determinant of $(I_N - \mc M_{+\ii})$, meaning we must apply the theorem to the negative matrix, $-\mc M_{+\ii}$. Multiplying an $N/2 \times N/2$ matrix block by $-1$ scales its determinant by exactly $(-1)^{N/2}$. Both diagonal-block determinants of $-\mc M_{+\ii}$ therefore have sign $(-1)^{N/2}$, and the theorem yields:
\begin{equation}
    (-1)^{N/2}\det(I_N-\mc M_{+\ii})\ge0.
    \label{eq:trs_boson_three_plus_det_sign}
\end{equation}

Finally, Eq.~\eqref{eq:trs_eta_propagation} dictates that $\Tr A_{\ell,+\ii}=0$. Using Eqs.~\eqref{eq:trs_boson_three_plus_reduced_weight} and~\eqref{eq:trs_boson_three_plus_det_sign}, the complete Boltzmann weight becomes:
\begin{equation}
    W=\ee^{\sum_\ell\xi_\ell}
    (-1)^{N/2}\det(I_N-\mc M_{+\ii})^{-1}>0.
    \label{eq:trs_boson_three_plus_weight}
\end{equation}
As explicitly shown, the determinant inherently carries a sign of $(-1)^{N/2}$, which perfectly cancels the $(-1)^{N/2}$ factor originating from the $N/2$ creation directions. This exact cancellation establishes the strict positivity of the Boltzmann weight, proving that the split-orthogonal symmetry class is sign-problem-free.

\subsection{Time-reversal periodic table}
\label{app:trs_counterexamples}

The Kramers, pairing, and split-orthogonal classes constitute the fundamental sign-problem-free entries of the periodic table. Any larger symmetry set guarantees positivity provided it contains the symmetries of at least one of these three classes. The remaining possibilities are the $\{+,+\}$ and $\{-\}$ classes, along with their subsets. The following counterexamples demonstrate why these classes fail to guarantee positivity.

For the $\{+,+\}$ class, consider a single bosonic mode described by
\begin{equation}
    \hat h=\kappa(\hat n_b+1/2),\qquad
    \Theta_1(\hat b)=\hat b,\qquad \Theta_2(\hat b)=\ii\hat b.
    \label{eq:trs_boson_pp_counter_h}
\end{equation}
Both symmetry operations are antiunitary (complex-conjugating numerical coefficients), preserve $\hat h$, square to $+I$, and mutually anticommute. For $\kappa<0$, the sum over particle occupations converges, yielding $W=\ee^{\kappa/2}/(1-\ee^\kappa)$. However, for $\kappa>0$, the corresponding regularized weight becomes
\begin{equation}
    W=-\frac{1}{2\sinh(\kappa/2)}<0.
    \label{eq:trs_boson_pp_counter_weight}
\end{equation}
Both time-reversal symmetries act on the same oscillator, which cannot be separated into two independent conjugate bosonic subsystems. Its weight therefore lacks the modulus-square structure $W=W_+\overline{W_+}$ that preserves positivity under analytic continuation in the pairing class. This single-mode counterexample also rules out any general positivity guarantee for weaker symmetry classes containing only one such TRS, or lacking time-reversal symmetry entirely.

For $\{-\}$, consider the complete three-slice family
\begin{align}
    \hat h_1&=-\alpha(\hat n_b+\hat n_c+1)
    +\frac{x}{2}(\hat b^{+2}+\hat b^2+\hat c^{+2}+\hat c^2),\notag\\
    \hat h_2&=\ii t(\hat b^+\hat c^+-\hat b\hat c),\notag\\
    \hat h_3&=\frac{\ii t}{2}(\hat b^{+2}-\hat b^2-\hat c^{+2}+\hat c^2),
    \qquad \alpha,x,t\in\mathbb R.
    \label{eq:trs_boson_single_minus_counterexample}
\end{align}
Every slice preserves $\Theta(\hat b)=\hat c$, $\Theta(\hat c)=-\hat b$, with $\Theta^2=-I$. For $\alpha=0$ and $x=\pi$, all three pairing directions are present when $t\ne0$, and the family has no time-reversal symmetry beyond $\Theta^2=-I$. The resulting Gaussian weight is
\begin{equation}
    W=-\frac{1}{2(1+\cos^2t)}<0,
    \qquad 0<t\le\frac{\pi}{2},
    \label{eq:trs_boson_single_minus_weight}
\end{equation}
so, for example, $W=-1/2$ at $t=\pi/2$. This gives a negative weight in the single-$\{-\}$ class.

Thus $\{-,-\}$, $\{+,-\}$, and $\{+,+,+\}$ are the only fundamental bosonic sign-free classes. No proper subset suffices, and every sign-free class with more symmetries contains one of them. Table~\ref{tab:trs_boson_fermion} compares this result with the fermionic classification~\cite{li2016prl}. The Kramers class and the $\{+,-\}$ class are sign-free for both statistics; we call the latter the pairing class for bosons and the Majorana class for fermions. For the split-orthogonal class, Eq.~\eqref{eq:trs_boson_three_plus_det_sign} supplies a determinant sign $(-1)^{N/2}$. In the bosonic reduced trace, the same $N/2$ creation directions each contribute a minus sign because their commutators have the opposite sign, as in Eq.~\eqref{eq:trs_boson_reversed_mode}. These two factors cancel in Eq.~\eqref{eq:trs_boson_three_plus_weight}, giving a non-negative weight. Fermionic weights have no corresponding bosonic commutator factor, so the split-orthogonal class is not sign-free for fermions.

\begin{table}[ht]
    \caption{\textbf{Time-reversal symmetry classes for sign-problem-free bosonic and fermionic AFQMC.} The Kramers class $\{-,-\}$ and the $\{+,-\}$ class are sign-free for both statistics; the latter is called the pairing class for bosons and the Majorana class for fermions. The split-orthogonal class $\{+,+,+\}$ is sign-free only for bosons. Each sign records $\Theta_a^2=\pm I$. Unlisted classes at a given number of symmetries admit negative weights. The fermionic entries follow Ref.~\cite{li2016prl}.}
    \label{tab:trs_boson_fermion}
    \vspace{8pt}
    {\fontencoding{T1}\sffamily
    \begin{tikzpicture}[x=0.82\linewidth,y=1pt,
        every node/.style={inner sep=0pt,outer sep=0pt,font=\sffamily\footnotesize,align=center},
        heading/.style={text=white,font=\sffamily\bfseries\footnotesize,align=center},
        fundamental/.style={text=TRSAccent,font=\sffamily\bfseries\footnotesize,align=center},
        splitclass/.style={text=orange!75!black,font=\sffamily\bfseries\footnotesize,align=center}]
        \begin{scope}
            \clip[rounded corners=2pt] (0,0) rectangle (1,-130);
            \fill[TRSHeader]    (0,0)          rectangle (1,-24);
            \fill[TRSRowLight]  (0,-24)        rectangle (1,-42);
            \fill[TRSRowShade]  (0,-42)        rectangle (1,-68);
            \fill[TRSRowLight]  (0,-68)        rectangle (1,-110);
            \fill[TRSRowShade]  (0,-110)       rectangle (1,-130);
            \node[heading] at (0.14,-12) {No. of TRS};
            \node[heading] at (0.45,-12) {Bosonic sign-free classes};
            \node[heading] at (0.78,-12) {Fermionic sign-free classes};
            \node at (0.14,-33) {0,1};
            \node at (0.45,-33) {None};
            \node at (0.78,-33) {None};
            \node at (0.14,-55) {2};
            \node[fundamental] at (0.45,-55) {\{\textminus,\textminus\} = Kramers class\\ \{+,\textminus\} = pairing class};
            \node[fundamental] at (0.78,-55) {\{\textminus,\textminus\} = Kramers class\\ \{+,\textminus\} = Majorana class};
            \node at (0.14,-89) {3};
            \node[splitclass] at (0.45,-75) {\{+,+,+\} = split-orthogonal class};
            \node at (0.45,-84) {\{+,+,\textminus\}};
            \node at (0.45,-93) {\{+,\textminus,\textminus\}};
            \node at (0.45,-102) {\{\textminus,\textminus,\textminus\}};
            \node at (0.78,-78) {\{+,+,\textminus\}};
            \node at (0.78,-89) {\{+,\textminus,\textminus\}};
            \node at (0.78,-100) {\{\textminus,\textminus,\textminus\}};
            \node at (0.14,-120) {$\geq$\,4};
            \node at (0.45,-120) {All};
            \node at (0.78,-120) {All};
        \end{scope}
    \end{tikzpicture}}
\end{table}

\section{Proof of the reflection-positivity criterion}
\label{app:rp_proof}

Reflection positivity makes the configuration weight non-negative by expressing the propagation as a sum of products between two conjugate bosonic sectors. The trace of each such product is a modulus square. We first establish this property and its preservation under successive evolution steps, then derive the generator condition and its Gaussian block form.

Let $\mc F_L$ and $\mc F_R$ be the Fock spaces generated by $\hat b_i$ and $\hat c_i$ ($i=1,\ldots,M$), so that $\mc F=\mc F_L\otimes\mc F_R$. The corresponding operator algebras are denoted by $\mathfrak A_L$ and $\mathfrak A_R$. Operators acting on different sectors commute. The antilinear reflection $\theta:\mathfrak A_L\to\mathfrak A_R$ maps the two flavors into one another and complex conjugates their coefficients,
\begin{equation}
    \theta(\ii)=-\ii,\qquad
    \theta(\hat b_i)=\eta_i\hat c_i,\qquad
    \theta(\hat b_i^+)=\overline{\eta_i}\hat c_i^+,
    \qquad |\eta_i|=1.
    \label{eq:theta_def_sm}
\end{equation}
It preserves operator products, $\theta(\hat R\hat S)=\theta(\hat R)\theta(\hat S)$. The phases $\eta_i$ allow a fixed phase choice for each reflected mode.

Consider sums of reflected products,
\begin{equation}
    \hat X=\sum_r \hat R_r\,\theta(\hat R_r),
    \qquad
    \hat R_r\in\mathfrak A_L.
    \label{eq:rp_positive_set_sm}
\end{equation}
The trace of each term factorizes over the two Fock spaces. In occupation-number bases with matching particle occupations, reflection complex conjugates the left-sector trace. Hence, whenever the traces exist,
\begin{equation}
    \Tr_{\mc F}\left[\hat R\theta(\hat R)\right]
    =
    \Tr_{\mc F_L}(\hat R)\,
    \Tr_{\mc F_R}[\theta(\hat R)]
    =
    \left|\Tr_{\mc F_L}(\hat R)\right|^2
    \ge0.
\end{equation}
The phases in Eq.~\eqref{eq:theta_def_sm} correspond to a unitary change of the right-sector basis and leave its trace unchanged. Thus every finite sum in Eq.~\eqref{eq:rp_positive_set_sm} has a non-negative trace. We call these sums reflection-positive.

To apply this observation to imaginary-time evolution, the reflected-product form must survive multiplication. Commuting the right-sector operator past the left-sector operator gives
\begin{equation}
    \hat R\theta(\hat R)\,\hat S\theta(\hat S)
    =
    \hat R\hat S\,\theta(\hat R\hat S).
    \label{eq:rp_multiplication_sm}
\end{equation}
Expanding a product of sums therefore gives another sum of reflected products. If each slice has a reflection-positive propagator apart from a real scalar factor, $\ee^{\hat h_\ell(\phi_\ell)}=\ee^{\xi_\ell(\phi_\ell)}\hat X_\ell$ with $\xi_\ell\in\mathbb R$, the full product has the same form,
\begin{equation}
    \prod_{\ell=1}^{L}\ee^{\hat h_\ell(\phi_\ell)}
    =
    \exp\!\left[\sum_\ell\xi_\ell(\phi_\ell)\right]\hat X.
\end{equation}
Its trace is consequently non-negative,
\begin{equation}
    W(\phi)=\Tr\prod_{\ell=1}^{L}\ee^{\hat h_\ell(\phi_\ell)}\ge0,
\end{equation}
whenever the bare trace converges.

We next show how the generator condition produces this reflected-product structure. For one slice, write
\begin{equation}
    \hat h=
    \hat q+\theta(\hat q)
    +\sum_{\mu,\nu}C_{\mu\nu}\hat X_\mu\theta(\hat X_\nu)
    +\xi,
    \qquad
    C=C^\hc\succeq0,\quad \xi\in\mathbb R,
    \label{eq:rp_generator_sm}
\end{equation}
where $\hat q,\hat X_\mu\in\mathfrak A_L$. The terms $\hat q$ and $\theta(\hat q)$ describe evolution within the two sectors, while the matrix $C$ specifies their coupling. Positive semidefiniteness allows this coupling to be resolved into a sum of reflected products. Factor $C_{\mu\nu}=\sum_a L_{a\mu}\overline{L_{a\nu}}$ and define $\hat Y_a=\sum_\mu L_{a\mu}\hat X_\mu$. Antilinearity then gives
\begin{equation}
    \sum_{\mu,\nu}C_{\mu\nu}\hat X_\mu\theta(\hat X_\nu)
    =\sum_a\hat Y_a\theta(\hat Y_a).
\end{equation}

Evolution within the two sectors already has the required form, since $\hat q$ commutes with $\theta(\hat q)$,
\begin{equation}
    \ee^{s[\hat q+\theta(\hat q)]}
    =\ee^{s\hat q}\theta(\ee^{s\hat q}),\qquad s\ge0.
\end{equation}
For the coupling terms, expand the exponential. Commutation between the sectors turns every power into a reflected product,
\begin{equation}
    \ee^{s\hat Y_a\theta(\hat Y_a)}
    =\sum_{m=0}^{\infty}\frac{s^m}{m!}\,
    \hat Y_a^m\theta(\hat Y_a^m),\qquad s\ge0.
\end{equation}
Every coefficient is non-negative and can be absorbed into the two reflected factors through its square root. Combining the within-sector and coupling terms through the Lie-Trotter product formula~\cite{trotter1959product} gives
\begin{equation}
    \ee^{\hat h}
    =\ee^\xi\lim_{n\to\infty}
    \left[
    \ee^{[\hat q+\theta(\hat q)]/n}
    \prod_a\ee^{\hat Y_a\theta(\hat Y_a)/n}
    \right]^n.
\end{equation}
Each factor is built from reflected products, and Eq.~\eqref{eq:rp_multiplication_sm} preserves their form throughout the multiplication. The exponential series and Trotter limit therefore establish Eq.~\eqref{eq:rp_generator_sm} as a sufficient condition for non-negative weights in the trace-convergent regime.

The Gaussian form makes this criterion especially transparent. Choose the reflection convention $\eta_i=1$ and collect the operators in the left sector as
\begin{equation}
    \hat{\bb z}=(\hat{\bb b},\hat{\bb b}^+)^\trans,
    \qquad
    \theta(\hat{\bb z})=(\hat{\bb c},\hat{\bb c}^+)^\trans.
\end{equation}
A quadratic term acting within the left sector has the form $\hat q=\hat{\bb z}^\trans D\hat{\bb z}$; its reflected partner acts within the right sector with the conjugate matrix $\overline D$. The terms that couple the two sectors are collected in a matrix $F$. Grouping each flavor with its creation operators gives
\begin{equation}
    \hat\chi=(\hat{\bb z},\theta(\hat{\bb z}))^\trans
    = (\hat{\bb b},\hat{\bb b}^+,\hat{\bb c},\hat{\bb c}^+)^\trans,
\end{equation}
and the slice generator takes the block form
\begin{equation}
    \hat h=
    \hat\chi^\trans
    \begin{pmatrix}
        D & F\\
        F^\trans & \overline D
    \end{pmatrix}
    \hat\chi+\xi,
    \qquad
    F=F^\hc\succeq0,\quad \xi\in\mathbb R.
    \label{eq:rp_block_sm}
\end{equation}
The diagonal blocks describe the conjugate evolution within the two sectors, while $F$ describes their coupling. The single condition $F\succeq0$ is therefore the Gaussian form of the reflection-positive generator criterion. This is Eq.~\eqref{eq:rp_block_form} in the main text. The flavor-grouped order used here differs from the conventional Nambu order $(\hat{\bb b},\hat{\bb c},\hat{\bb b}^+,\hat{\bb c}^+)^\trans$ used elsewhere in the SM only by a fixed permutation of components. A reflection with phases $\eta_i$ is incorporated by the corresponding phase factors in $\theta(\hat{\bb z})$.
The relation to the TRS criterion is direct. For a number-conserving Hamiltonian, the absence of anomalous pairing terms sets $F=0$ in the RP block form. The remaining flavor blocks are complex conjugates. In the flavor space spanned by $(\hat b_i,\hat c_i)^\trans$, this structure is invariant under the TRS $\Theta=\ii\sigma_y K$, which satisfies $\Theta^2=-I$ and realizes the Kramers class; number-conserving RP is thus contained in the Kramers TRS criterion. Allowing pairing admits $F\succeq0$ and extends RP beyond the TRS classes. In the triangular-lattice paired benchmark, the antiunitary maps relating the two flavors commute and hence do not form any sign-free class in Table~\ref{tab:trs_periodic}, whereas the reflected anomalous coupling remains positive. RP is thus the mechanism that protects its weight.

Successive RP evolution steps combine into another RP propagator, as shown in Eq.~\eqref{eq:rp_multiplication_sm}. Such a set, closed under associative composition without requiring inverse propagators to belong to it, is called a semigroup. For fermions, Majorana RP within the reflection-symmetric sector can be expressed as the contraction condition $A^\hc\eta+\eta A\preceq0$ for a fixed indefinite Hermitian metric $\eta$~\cite{wei2024prb}. For bosons, no fixed Hermitian metric $\eta$ makes this contraction condition equivalent to the full algebraic RP condition $F\succeq0$. To prove this by contradiction, suppose such a metric existed. At $D=0,F=I$ in Eq.~\eqref{eq:rp_block_sm}, the Nambu generator satisfies $A^\hc=-A$, hence $\Tr(A^\hc\eta+\eta A)=0$. The contraction inequality then forces $A^\hc\eta+\eta A=0$, since nonpositive eigenvalues with zero sum must all vanish. Reversing the coupling to $F=-I$ sends $A$ to $-A$ and preserves this equality. The same metric would therefore admit a coupling that violates $F\succeq0$, contradicting the assumed equivalence. Thus the bosonic RP cone defines a semigroup that admits no equivalent description as a contraction semigroup with a fixed Hermitian metric. The bosonic RP cone revealed here may provide key insight for the future development of a broader semigroup theory of sign-free quantum Monte Carlo beyond the fixed-metric contraction framework of Ref.~\cite{wei2024prb}.

\section{Benchmark models and numerical details}
\label{app:benchmark_details}

We use the triangular lattice with primitive vectors $\bb a_1=(1,0)$ and $\bb a_2=(1/2,\sqrt3/2)$, keeping $N_s=9$ sites with periodic boundary conditions. The general two-flavor benchmark Hamiltonian is
\begin{equation}
\begin{aligned}
    \hat H={}&t\sum_{\langle ij\rangle}
    \left(\hat b_i^+\hat b_j+\hat b_j^+\hat b_i
    +\hat c_i^+\hat c_j+\hat c_j^+\hat c_i\right)
    -\sum_i\left(\Delta\hat b_i^+\hat c_i^+ +\Delta^*\hat b_i\hat c_i\right)\\
    &+U_1\sum_i(\hat n_{b,i}-\hat n_{c,i})^2
    +U_2\sum_i(\hat n_{b,i}+\hat n_{c,i})^2,
\end{aligned}
\label{eq:benchmark_general}
\end{equation}
where $t>0$, $\hat n_{b,i}=\hat b_i^+\hat b_i$, and $\hat n_{c,i}=\hat c_i^+\hat c_i$. We sample $\hat H-\mu\hat N$ in the grand-canonical ensemble, with $\hat N=\sum_i(\hat n_{b,i}+\hat n_{c,i})$. The two density channels are decoupled by the HS transformations in Eqs.~\eqref{eq:benchmark_hs_total} and~\eqref{eq:benchmark_hs_relative}.

For all benchmarks, we measure the density $\rho=\langle\hat N\rangle/N_s$, the energy $E=\langle\hat H\rangle$ excluding the chemical-potential term, and the superfluid and density-wave structure factors at $K=(4\pi/3,0)$. We plot the energy as $-E$,
\begin{align}
    S_{\rm SF}(K)&=N_s^{-2}\sum_{ij}\ee^{\ii K\cdot(r_i-r_j)}
    \langle\hat b_i^+\hat b_j+\hat c_i^+\hat c_j\rangle,\\
    S_{\rm DW}(K)&=N_s^{-2}\sum_{ij}\ee^{\ii K\cdot(r_i-r_j)}
    \langle(\hat n_{b,i}+\hat n_{c,i})(\hat n_{b,j}+\hat n_{c,j})\rangle.
\end{align}
Here $r_i$ is the position of site $i$.

\begin{figure}[tb]
    \centering
    \includegraphics[width=\linewidth]{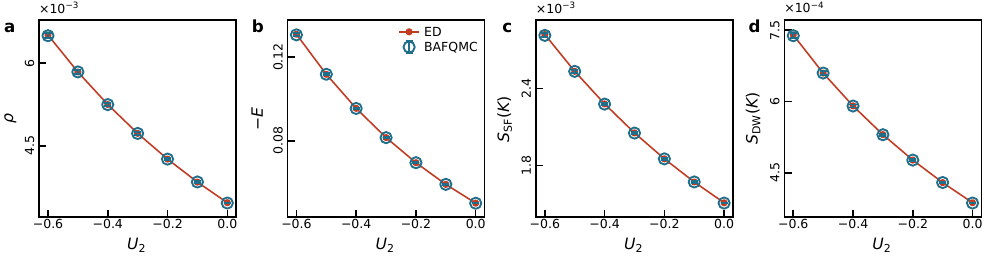}
    \caption{\textbf{Attractive-density benchmark.} Results for Eq.~\eqref{eq:benchmark_general} with $\Delta=0$ on the periodic $3\times3$ triangular lattice at $t=1$, $U_1=1$, $\mu=-7$, and $\beta=1$. The panels show $\rho$, $-E$, $S_{\rm SF}(K)$, and $S_{\rm DW}(K)$ as $U_2$ varies. Red points denote finite-occupation reference values, and blue open circles denote BAFQMC means with standard-error bars.}
    \label{fig:benchmark_attractive}
\end{figure}

We first consider the number-conserving benchmarks, with $\Delta=0$. For each real auxiliary-field configuration, the two flavors experience complex-conjugate potentials, $v_{b,i\ell}=\alpha_2\phi_{i\ell}+\ii\alpha_1\varphi_{i\ell}$ and $v_{c,i\ell}=\overline{v_{b,i\ell}}$, while their hopping and chemical-potential terms are identical. Every slice is therefore invariant under the antiunitary map $\Theta=\ii\sigma_yK$, which exchanges $b$ and $c$ and satisfies $\Theta^2=-I$. Together with particle-number conservation, this realizes the Kramers class $\{-,-\}$ of Sec.~\ref{app:trs_kramers}. Physically, the fixed-field $b$ and $c$ evolutions are exact complex conjugates. Their configuration weights and equal-time Green functions obey $W_c=\overline{W_b}$ and $G_c=\overline{G_b}$, so the full weight is simply $W=W_bW_c=|W_b|^2\ge0$. We evaluate only the $b$-flavor propagation and reconstruct the $c$-flavor weight factor and Green function by complex conjugation. The repulsive relative-density benchmark in Fig.~\ref{fig:benchmark} uses $U_2=0$, $t=1$, $\beta=4$, $\Delta\tau=0.01$, and $\mu=-3.5$, with $U_1$ varied. The attractive total-density benchmark in Fig.~\ref{fig:benchmark_attractive} uses $U_1=1$, $t=1$, $\mu=-7$, $\beta=1$, and $\Delta\tau=0.01$, with $U_2\le0$ varied. For $U_2<0$, the bare trace of an individual auxiliary-field configuration can diverge; the configuration weights sampled in this scan are the regularized weights defined in Sec.~\ref{app:regularization}. In both scans, the BAFQMC results agree with finite-temperature exact diagonalization within statistical errors for all observables.

The paired benchmark in Fig.~\ref{fig:benchmark} uses $U_1=1$, $U_2=0$, $t=1$, $\beta=4$, $\Delta\tau=0.01$, and $\mu=-5$, with the pairing amplitude varied. The BAFQMC observables show quantitative agreement with finite-temperature exact diagonalization. Reflection positivity is obtained by reflecting between the two flavors with $\theta(\hat b_i)=\eta\hat c_i$ and $\theta(\hat b_i^+)=\overline\eta\hat c_i^+$, where $\eta=\Delta^*/|\Delta|$ for $\Delta\ne0$ (and $\eta=1$ when $\Delta=0$). The pairing part of each slice is then $\Delta\tau|\Delta|\sum_i[\hat b_i^+\theta(\hat b_i^+)+\hat b_i\theta(\hat b_i)]$. Thus every pair term is a product of a $b$-sector operator and its reflected $c$-sector partner with a non-negative coefficient. The hopping, chemical-potential, and density-field terms remain two conjugate within-sector evolutions, while the cross-flavor block is $F=(\Delta\tau|\Delta|/2)I_{2N_s}\succeq0$. Each slice is therefore reflection positive, and successive slices preserve this structure; the convergent trace is a sum of modulus squares and is non-negative.

For the paired benchmark with $\Delta\ne0$, $U_1>0$, and $U_2=0$, we now prove that the imaginary-time evolution operator has a convergent trace for every auxiliary-field configuration. Let $\hat K_0$ denote the hopping, pairing, and chemical-potential terms in $\hat H-\mu\hat N$. The triangular-lattice dispersion is $\epsilon_{\bb k}=t|1+\ee^{\ii\bb k\cdot\bb a_1}+\ee^{\ii\bb k\cdot\bb a_2}|^2-3t\ge-3t$. Thus $\mu<-3t-|\Delta|$ makes all Bogoliubov energies $\omega_{\bb k}=\sqrt{(\epsilon_{\bb k}-\mu)^2-|\Delta|^2}$ positive, and the quadratic propagator $T=\ee^{-\Delta\tau\hat K_0}$ is trace class, with
\begin{equation}
    Z_0=\Tr\,\ee^{-\beta\hat K_0}
    =\prod_{\bb k}\frac{\ee^{\beta(\epsilon_{\bb k}-\mu-\omega_{\bb k})}}
    {(1-\ee^{-\beta\omega_{\bb k}})^2}<\infty.
    \label{eq:benchmark_free_partition}
\end{equation}
For real HS fields, $V_\ell=\exp[\ii\sqrt{2\Delta\tau U_1}\sum_i\varphi_{i\ell}(\hat n_{b,i}-\hat n_{c,i})]$ is unitary: in the occupation basis it only changes phases and preserves norms. Each time slice therefore contains the same positive thermal propagator $T$, with such a phase rotation inserted between successive propagators. Applying H\"older's inequality to $\hat U(\phi)=\prod_{\ell=1}^{L_\tau}TV_\ell$ gives
\begin{equation}
    \left|\Tr\hat U(\phi)\right|
    \le\prod_{\ell=1}^{L_\tau}
    \left[\Tr|TV_\ell|^{L_\tau}\right]^{1/L_\tau}
    =\Tr T^{L_\tau}=Z_0<\infty,
    \label{eq:benchmark_trace_bound}
\end{equation}
where $|X|=(X^\hc X)^{1/2}$. The equality uses $|TV_\ell|=V_\ell^\hc T V_\ell$, unitarity of $V_\ell$, and $L_\tau\Delta\tau=\beta$. Thus the thermal suppression of large occupations controls the full evolution even with arbitrary HS phase rotations between slices. The bound is independent of the auxiliary fields and proves trace convergence for every configuration. Together with reflection positivity, it gives $0\le W(\phi)\le Z_0$. This bound also excludes bosonic poles. In Eq.~\eqref{eq:paired_trace_formula_appendix}, the scalar prefactor accounts for normal ordering in the Nambu representation and is finite and nonzero. A zero of the determinant would therefore make the weight diverge, contradicting Eq.~\eqref{eq:benchmark_trace_bound}. Hence $\det(I_{2N}-\mc M)\ne0$ for every auxiliary-field configuration. Under this stability condition, BAFQMC encounters neither a bosonic pole nor the infinite-variance problem.

With the positivity and finiteness of every configuration established, we now give the simulation parameters and the details of the exact-diagonalization reference. Each parameter point uses $\Delta\tau=0.01$, giving $L_\tau=400$ slices at $\beta=4$ and $L_\tau=100$ at $\beta=1$. We perform $500$ warmup sweeps, followed by $10^5$ measurement sweeps with one measurement bin per sweep. Statistical errors are standard errors over ten blocks of $10^4$ bins. Local field proposals are uniform shifts in $[-0.75,0.75]$ for number-conserving runs and $[-0.5,0.5]$ for paired runs. We stabilize matrix products every $N_{\rm wrap}=10$ slices using Secs.~\ref{app:stabilization_number} and~\ref{app:stabilization_paired}. The BAFQMC, ED, and benchmark implementations are available in Ref.~\cite{bafqmc_code}.

Finite-temperature ED provides the reference values in Figs.~\ref{fig:benchmark} and~\ref{fig:benchmark_attractive}. We construct the Hamiltonian matrices and diagonalize them with QuSpin~\cite{weinbergBukov2019quspin}, while the finite-temperature sums and observable estimators are implemented in our own wrapper. At $U_1=U_2=0$ and $\Delta=0$, we use the exact free-boson result. For the repulsive $U_1$ scan, we diagonalize separately in each total-particle-number sector and include sectors through $N_{\rm tot}=10$; the last-shell contributions to the plotted observables are below $10^{-4}$ relative to their accumulated values. For the attractive $U_2$ scan, we include six complete particle shells, up to the configured cutoff $N_{\rm tot}=8$. We accept these low-density reference values when the last-shell contribution to each observable is below $10^{-5}$ relative to its accumulated value and the total density is below $0.05$. For finite pairing, the ED basis imposes the local cutoffs $n_{b,i},n_{c,i}\le3$ and the total-particle cutoff $N_{\rm tot}\le4$, giving 7297 states at the largest pairing amplitude. At this endpoint, the thermal weight on the total-particle cutoff boundary is $1.1\times10^{-4}$, while that on the local-occupation boundary is below $4\times10^{-14}$.

\section{More sign-problem-free models as examples and applications}
\label{app:trs_models}

In this section, we present more bosonic models of physical interest that can be simulated with sign-problem-free BAFQMC. These models have so far remained inaccessible to existing sign-free QMC methods. We first give a Kramers-class model of interacting $p$-orbital bosons with stripe superfluidity, then a pairing-class model coupling opposite chiral bosonic bands on the kagome lattice, and finally a split-orthogonal model of flavor-singlet odd-parity pairing on the frustrated triangular lattice. The last construction has a special status: the split-orthogonal class is the unique fundamental TRS class that guarantees sign-free odd-parity bosonic pairing, including $p$-wave pairing. The proof in Sec.~\ref{app:trs_two_flavor_odd_pairing} establishes this uniqueness by excluding the Kramers and pairing classes for this odd-parity pairing structure.

\subsection{Kramers class: stripe superfluidity in frustrated \texorpdfstring{$p$-band bosons}{p-band bosons}}
\label{sec:trs_orbital_hund}

Orbital bosons support spontaneous circulation, finite-momentum condensation, and stripe order without an applied gauge field. In $p$ bands, local interactions select angular-momentum-carrying $p_x\pm\ii p_y$ combinations through the orbital Hund channel, while directional tunneling and frustrated lattice geometry organize these circulating states into nonzero-momentum condensates and stripes~\cite{liuWu2006orbital,wuLiuMoore2006stripe,wu2009unconventional}. Interactions can also couple a spin doublet to orbital angular momentum in $p$-orbital bosons, producing an intertwined superfluid with Dirac excitations~\cite{liYuanHemmerichLi2018prl}. The model below combines these spin and orbital degrees of freedom with frustrated triangular-lattice hopping and a Kramers-protected sign-free weight. Quantum stripe order has also been observed in a triangular optical lattice~\cite{wangLuoLiu2023stripe}.

Place two internal states $\sigma=\uparrow,\downarrow$ in the $p_x,p_y$ orbitals of each triangular-lattice site. Write $\hat{\bb p}_{i\sigma}=(\hat p_{ix\sigma},\hat p_{iy\sigma})$ and define the occupation and orbital angular momentum, in units of $\hbar$, by
\begin{equation}
    \hat n_{i\sigma}=\hat p_{ix\sigma}^{+}\hat p_{ix\sigma}
       +\hat p_{iy\sigma}^{+}\hat p_{iy\sigma},
    \qquad
    \hat L_{i\sigma}=-\ii\bigl(
       \hat p_{ix\sigma}^{+}\hat p_{iy\sigma}
       -\hat p_{iy\sigma}^{+}\hat p_{ix\sigma}\bigr).
    \label{eq:trs_minimal_orbital_operators}
\end{equation}
The Hamiltonian $\hat H=\hat H_0+\hat H_I$ is
\begin{align}
    \hat H_0={}&t\sum_{\langle i,j\rangle,\sigma}
       \Bigl[(\bb e_{ij}\cdot\hat{\bb p}_{i\sigma}^{+})
             (\bb e_{ij}\cdot\hat{\bb p}_{j\sigma})
             +\mathrm{h.c.}\Bigr]
       -\mu\sum_{i,\sigma}\hat n_{i\sigma},
       \label{eq:trs_minimal_pband_H0}\\
    \hat H_I={}&\frac U2\sum_i
       \left[(\hat n_{i\uparrow}-\hat n_{i\downarrow})^2
       -\frac13(\hat L_{i\uparrow}-\hat L_{i\downarrow})^2\right],
       \qquad t,U>0.
       \label{eq:trs_minimal_pband_HI}
\end{align}
Here $\bb e_{ij}$ is the unit vector along a nearest-neighbor bond, and each bond is counted once. The hopping retains only the dominant longitudinal overlap; its positive sign follows from the odd parity of the $p$ orbitals~\cite{wuLiuMoore2006stripe}.

Expanding Eq.~\eqref{eq:trs_minimal_pband_HI} gives the standard single-component $p$-orbital interaction $\frac U2\sum_\sigma(\hat n_{i\sigma}^2-\hat L_{i\sigma}^2/3)$~\cite{liuWu2006orbital,wu2009unconventional}, together with $-U\hat n_{i\uparrow}\hat n_{i\downarrow}+(U/3)\hat L_{i\uparrow}\hat L_{i\downarrow}$ between components. Linear normal-ordering terms can be absorbed into $\mu$. This is a local effective density-and-angular-momentum model: it retains the single-component contact Hund interaction and describes intercomponent coupling through these two channels. The density square penalizes population imbalance, whereas the negative angular-momentum square favors locally circulating orbitals with opposite circulation in the two components.

On each internal-state doublet, take $\Theta=\ii\sigma_y\mc K$, with $\Theta^2=-I$. The hopping is real and component independent. Under $\Theta$, the density difference is odd, while the angular-momentum difference is even, since $\hat L_{i\uparrow}\mapsto-\hat L_{i\downarrow}$. The two commuting onsite squares have the exact HS decomposition
\begin{equation}
    \ee^{-\Delta\tau\hat H_{I,i}}
    =\int\frac{\dd\phi\,\dd\eta}{2\pi}\,
       \ee^{-(\phi^2+\eta^2)/2}
       \exp\Bigl[\ii\sqrt{U\Delta\tau}\,\phi
                 (\hat n_{i\uparrow}-\hat n_{i\downarrow})
                 +\sqrt{U\Delta\tau/3}\,\eta
                 (\hat L_{i\uparrow}-\hat L_{i\downarrow})\Bigr].
    \label{eq:trs_minimal_orbital_HS}
\end{equation}
Every field value preserves $\Theta$: the odd density channel has an imaginary coupling, and the even angular-momentum channel has a real coupling. Because particle number is conserved, the two component propagators remain complex conjugates, $\mc B_{\downarrow}=\mc B_{\uparrow}^{*}$, giving $W=|\det(I-\mc B_{\uparrow})|^{-2}>0$. Thus only two real auxiliary fields per site are needed despite the frustrated orbital hopping.

The band structure connects local circulation to orbital stripes. The lower band has three symmetry-related minima at the Brillouin-zone edge centers $M_1,M_2,M_3$~\cite{wuLiuMoore2006stripe}. A coherent equal-weight superposition of two minima with relative phase $\pi/2$, together with its complex conjugate in the other component, gives a representative counter-rotating stripe condensate. At density $\rho$ per component, choosing horizontal stripes yields
\begin{equation}
    \langle\hat n_{i\uparrow}\rangle
       =\langle\hat n_{i\downarrow}\rangle=\rho,
    \qquad
    \langle\hat L_{i\uparrow}\rangle
       =-\langle\hat L_{i\downarrow}\rangle
       =\frac{\sqrt3}{2}\rho\,(-1)^m,
    \label{eq:trs_minimal_orbital_stripe}
\end{equation}
where $m$ labels horizontal lattice rows; the factor $(-1)^m$ is the interference pattern from the momentum difference of the two selected $M$ points. The site density is uniform, but the circulation reverses between adjacent rows and between components. Relative to a noncirculating single-$M$ condensate at the same filling, this coherent state lowers the interaction energy per site by $U\rho^2/2$ while retaining the band-minimum kinetic energy. The linear contributions from the operator squares are identical in this comparison. The Hund channel therefore selects a stripe superfluid with counter-rotating orbital currents from the degenerate band minima.

BAFQMC can follow this competition across the phase diagram. Single-particle correlations measure superfluid coherence, while correlations of $\hat L_{i\uparrow}-\hat L_{i\downarrow}$ reveal the orbital stripes even when the total angular momentum cancels. These observables distinguish a stripe superfluid, where phase coherence and staggered circulation coexist, from an orbital-stripe phase without a single-particle condensate, and they resolve how the three symmetry-related $M$ minima select competing domains.

\subsection{Pairing class: pairing-induced chiral bands}

Pair creation with spatially varying phases provides a route to topological bosonic matter. On the kagome lattice, this mechanism yields Chern Bogoliubov bands and chiral edge modes in nonlinear bosonic systems and parametrically coupled photonic cavities~\cite{peanoHoudeBrendel2016squeezing,bardynKarzigRefaelLiew2016bogoliubov}. The same pairing structure supports topologically protected quantum-limited amplification and chiral transport~\cite{peanoHoudeMarquardt2016topological}, and has been realized experimentally in a bosonic Kitaev chain with chiral amplification~\cite{slimWanjuraBrunelli2024bosonKitaev}. We use this phase-winding mechanism to construct an interacting model in which a local interaction can mix the two conjugate chiral sectors.

Place bosons $\hat b_i$ and $\hat c_i$ on the sites of a kagome lattice, with sublattices $A,B,C$ labelled counterclockwise around an upward triangle. The pairing phases are $(\varphi_A,\varphi_B,\varphi_C)=(0,2\pi/3,4\pi/3)$, and $\varphi_i$ is the phase of the sublattice containing site $i$. Consider $\hat H=\hat H_0+\hat H_I$, with
\begin{align}
    \hat H_0={}&\omega_0\sum_i(\hat n_{b,i}+\hat n_{c,i})
       -t\sum_{\langle i,j\rangle}
       (\hat b_i^+\hat b_j+\hat b_j^+\hat b_i
        +\hat c_i^+\hat c_j+\hat c_j^+\hat c_i)\notag\\
       &-\frac{\Delta_{\rm s}}{2}\sum_i\left[
         \ee^{\ii\varphi_i}(\hat b_i^+)^2
        +\ee^{-\ii\varphi_i}(\hat c_i^+)^2
        +\ee^{-\ii\varphi_i}\hat b_i^2
        +\ee^{\ii\varphi_i}\hat c_i^2\right],
    \label{eq:trs_conjugate_BdG_H0}\\
    \hat H_I={}&U\sum_i(\hat n_{b,i}-\hat n_{c,i})^2,
    \qquad U>0,
    \label{eq:trs_conjugate_BdG_HI}
\end{align}
Here $\omega_0>0$ is the onsite energy, $t>0$ is the hopping amplitude, and $\Delta_{\rm s}>0$ is the pair-creation amplitude. Each nearest-neighbor bond is counted once. The onsite pairing creates two bosons of the same flavor, with a phase that winds in opposite directions for $b$ and $c$ around a triangle. At $U=0$, each flavor realizes the kagome squeezing model of Ref.~\cite{peanoHoudeBrendel2016squeezing} with onsite pair creation. The phase pattern opens gaps and produces Bogoliubov bands with nonzero Chern numbers. Complex conjugation reverses those Chern numbers, so the two flavors carry oppositely propagating edge modes. Chirality is thus resolved by flavor in this time-reversal-invariant two-component system.

A sufficient stability condition is $\omega_0>4t+\Delta_{\rm s}$, since the kagome coordination number is four. It makes the Nambu quadratic form positive definite, while $\hat H_I\ge0$ preserves stability. On a finite lattice, $\ee^{-\Delta\tau\hat H_0}$ then has a convergent trace. The auxiliary-field factors below are unitary, so their products with these quadratic factors also have finite traces.

On the flavor doublet $(\hat b_i,\hat c_i)^\trans$, the two time reversals $\Theta_1=\sigma_x\mc K$ and $\Theta_2=\ii\sigma_y\mc K$ obey $\Theta_1^2=+I$, $\Theta_2^2=-I$, and $\Theta_1\Theta_2=-\Theta_2\Theta_1$. Both preserve $\hat H_0$ and reverse the number imbalance, $\hat n_{b,i}-\hat n_{c,i}\mapsto-(\hat n_{b,i}-\hat n_{c,i})$.
The continuous Hubbard--Stratonovich identity
\begin{equation}
    \ee^{-\Delta\tau U(\hat n_{b,i}-\hat n_{c,i})^2}
    =\frac{1}{\sqrt{2\pi}}
    \int_{-\infty}^{\infty}d\phi_i\,\ee^{-\phi_i^2/2}
    \ee^{\ii\sqrt{2\Delta\tau U}\,\phi_i
      (\hat n_{b,i}-\hat n_{c,i})}
    \label{eq:trs_conjugate_BdG_relative_density_HS}
\end{equation}
therefore preserves both time reversals for every real value of $\phi_i$. Complex conjugation changes the sign of $\ii$, while the exchange of $b$ and $c$ supplies the second sign change. At fixed auxiliary fields, the \(b\) and \(c\) evolutions remain separate and are exact complex conjugates. Their many-body traces obey
\begin{equation}
    W(\phi)=W_b(\phi)\overline{W_b(\phi)}
    =|W_b(\phi)|^2>0.
    \label{eq:trs_conjugate_BdG_weight}
\end{equation}

The onsite interaction links this chiral pairing structure to a distinct many-body ordering problem. Expanding $U(\hat n_{b,i}-\hat n_{c,i})^2$ gives an interflavor attraction $-2U\hat n_{b,i}\hat n_{c,i}$ together with repulsion within each flavor. The interaction favors equal local populations. Starting from equal populations, adding a single boson costs an imbalance energy $U$, whereas adding one $b$ boson and one $c$ boson together leaves the imbalance unchanged. Interflavor pairs can therefore fluctuate without this energy cost even when single-boson fluctuations are suppressed, motivating coherent pair order $\langle\hat b_i\hat c_i\rangle\ne0$. Such order breaks the separate flavor parities preserved by the imposed same-flavor pairing. A nonzero pair amplitude acts at quadratic level as an interflavor term proportional to $\hat b_i\hat c_i+\mathrm{h.c.}$ and hence mixes the two conjugate BdG sectors. Because these sectors carry opposite Chern numbers and counterpropagating edge modes, the pair order directly hybridizes the edge channels and can open an interaction-induced edge gap. The same mechanism can drive a bulk gap closing and reopening, giving a concrete interaction-driven transition between a chiral-band regime and an interflavor-paired regime. This is the many-body extension of the phase-winding mechanism used in topological photonic systems, including nonlinear photonic crystals and superconducting cavity arrays~\cite{peanoHoudeBrendel2016squeezing,bardynKarzigRefaelLiew2016bogoliubov,peanoHoudeMarquardt2016topological}. BAFQMC can resolve this transition, its edge-mode reconstruction, and the associated pair correlations without a sign problem.

\subsection{Split-orthogonal class: flavor-singlet odd-parity pairing}
\label{app:trs_two_flavor_odd_pairing}

Odd-parity pairing links unconventional Bose gases and frustrated quantum magnetism through an antisymmetric flavor channel. In resonant mixtures it supports finite-momentum superfluidity~\cite{radzihovskyChoi2009pwave,choiRadzihovsky2011finite}; in rotating two-component gases it yields exact paired states~\cite{papenbrockReimannKavoulakis2012pwave}; and in Schwinger-boson theories it forms the spin-singlet bonds from which spin-liquid states are built~\cite{readSachdev1991frustrated,sachdev1992kagome}. Across these settings, Bose exchange symmetry requires an odd spatial pair to be antisymmetric in flavor. For two flavors this gives the unique channel
\begin{equation}
    \hat P_{ij}^+=\hat b_i^+\hat c_j^+-\hat b_j^+\hat c_i^+,
    \qquad \hat P_{ij}=\hat c_j\hat b_i-\hat c_i\hat b_j,
    \qquad \hat P_{ji}^+=-\hat P_{ij}^+.
    \label{eq:trs_odd_pair}
\end{equation}
Bosonic commutation relations forbid same-flavor odd-parity pairing, while allowing the flavor-singlet pair in Eq.~\eqref{eq:trs_odd_pair}.

We first determine its allowed TRS algebra. Let a time reversal act by the same unitary matrix $U$ on $(\hat b_i,\hat c_i)^\trans$ at every site, without mixing creation and annihilation operators. Since
\begin{equation}
    \hat P_{ij}^+=(\hat b_i^+,\hat c_i^+)\ii\sigma_y
                    \begin{pmatrix}\hat b_j^+\\\hat c_j^+\end{pmatrix},
    \qquad
    U^\trans(\ii\sigma_y)U=(\det U)\ii\sigma_y,
    \label{eq:trs_two_by_two_determinant_identity}
\end{equation}
and $\ii\sigma_y$ is real, antiunitarity and Hermitian conjugation give
\begin{equation}
    \Theta(\hat P_{ij}^+)=(\det U)^*\hat P_{ij}^+,
    \qquad
    \Theta(\hat P_{ij})=(\det U)\hat P_{ij}.
    \label{eq:trs_odd_pair_determinant_transformation}
\end{equation}
For a nonzero complex bond amplitude, invariance requires $\Delta_{ij}^*(\det U)^*=\Delta_{ij}$. Thus all nonzero amplitudes preserved by a uniform internal time reversal share the same phase up to signs. Choosing that common phase to be real, preservation of the pair field requires $\det U=1$.

A time-reversal symmetry with $\Theta^2=-I$ additionally satisfies $\overline U U=-I_2$. Unitarity then gives $U^\trans=-U$, and the determinant-one condition restricts it to $U=\pm\ii\sigma_y$. The corresponding time reversals $\pm\Theta_0$, with $\Theta_0=\ii\sigma_y\mc K$, differ only by a sign and cannot supply the two anticommuting time reversals required by the Kramers class. A time reversal $\Theta=U\mc K$ with $\Theta^2=+I$ must also satisfy $\det U=1$ to preserve the pair field. Together with unitarity, this gives $U(\ii\sigma_y)=(\ii\sigma_y)\overline U$, and hence $\Theta\Theta_0=\Theta_0\Theta$. It therefore cannot supply the anticommuting partner of $\Theta_0$ required by the pairing class.

The three time-reversal symmetries with $\Theta_a^2=+I$,
\begin{equation}
    \Theta_1=\mc K,\qquad
    \Theta_2=\ii\sigma_x\mc K,\qquad
    \Theta_3=\ii\sigma_z\mc K
    \label{eq:trs_odd_pair_three_plus_TR}
\end{equation}
all preserve the pair because their matrices have determinant one. The Pauli algebra gives pairwise anticommutation. Hence the odd-parity field realizes the split-orthogonal class $\{+,+,+\}$, the only fundamental sign-free class compatible with this field under uniform internal time reversal. This conclusion follows from the flavor algebra and is independent of lattice geometry.

A concrete interacting realization uses two flavors on the triangular lattice. Broad interspecies $p$-wave resonances have been observed in Rb mixtures~\cite{dongCuiShen2016pwave}, and ultracold heteronuclear $p$-wave molecules have been created from Na and Rb atoms~\cite{jiaGuoHuangWang2026pwave}. A coherent molecular field supplies an atomic pair field with the structure in Eq.~\eqref{eq:trs_odd_pair}. Consider real $t>0$ and $\Delta_{\rm p}>0$,
\begin{align}
    \hat H={}&\hat H_0+\hat H_I,
    \label{eq:trs_three_plus_model}\\
    \hat H_0={}&-\mu\sum_i(\hat n_{b,i}+\hat n_{c,i})
      +t\sum_{\langle i,j\rangle}
       (\hat b_i^+\hat b_j+\hat b_j^+\hat b_i
        +\hat c_i^+\hat c_j+\hat c_j^+\hat c_i)
      -\Delta_{\rm p}\sum_{\langle i,j\rangle}(\hat P_{ij}^++\hat P_{ij}),
    \label{eq:trs_three_plus_model_H0}\\
    \hat H_I={}&g_1\sum_{\langle i,j\rangle}\hat J_{ij}^2
       +g_2\sum_{\langle i,j\rangle}
          [-\ii(\hat P_{ij}^+-\hat P_{ij})]^2,
    \qquad g_1,g_2>0,
    \label{eq:trs_three_plus_model_HI}\\
    \hat J_{ij}={}&-\ii(\hat b_i^+\hat b_j-\hat b_j^+\hat b_i
                       +\hat c_i^+\hat c_j-\hat c_j^+\hat c_i).
    \label{eq:trs_three_plus_total_current}
\end{align}
Each bond is counted once, oriented along $\bb a_1=(1,0)$, $\bb a_2=(1/2,\sqrt3/2)$, or $\bb a_2-\bb a_1$. This fixes the signs of the odd-parity bond field. The current square combines density interactions, pair hopping, and coupling between the two flavor currents. The second square contains pair scattering and coherent creation or removal of two bond pairs. Both interactions preserve global flavor $SU(2)$, as does $\hat H_0$.

The quadratic part is invariant under all three maps in Eq.~\eqref{eq:trs_odd_pair_three_plus_TR}. Each map reverses $\hat J_{ij}$ and $-\ii(\hat P_{ij}^+-\hat P_{ij})$: complex conjugation reverses the explicit $\ii$, while the flavor transformations preserve the total hopping and the singlet pair. Decoupling the two bond-interaction factors separately gives
\begin{equation}
\begin{aligned}
    \ee^{-\Delta\tau g_1\hat J_{ij}^2}
    &=\int\frac{\dd\phi_{ij}}{\sqrt{2\pi}}\,
      \ee^{-\phi_{ij}^2/2}
      \ee^{\ii\sqrt{2\Delta\tau g_1}\,\phi_{ij}\hat J_{ij}},\\
    \ee^{-\Delta\tau g_2[-\ii(\hat P_{ij}^+-\hat P_{ij})]^2}
    &=\int\frac{\dd\varphi_{ij}}{\sqrt{2\pi}}\,
      \ee^{-\varphi_{ij}^2/2}
      \ee^{\ii\sqrt{2\Delta\tau g_2}\,\varphi_{ij}
        [-\ii(\hat P_{ij}^+-\hat P_{ij})]}.
    \label{eq:trs_three_plus_HS}
\end{aligned}
\end{equation}
Here $\phi_{ij}$ and $\varphi_{ij}$ are independent real fields, one for each bond and interaction channel. Each field value preserves all three time reversals: antiunitarity reverses $\ii$, and the corresponding Hermitian operator is odd under every symmetry, so the two sign changes cancel. Theorem~\ref{thm:trs} therefore guarantees positive configuration weights despite frustrated hopping and odd-parity pairing.

For a stable grand-canonical realization, choose $\mu$ so that the quadratic Bogoliubov energies are positive. For the equal-amplitude model in Eq.~\eqref{eq:trs_three_plus_model_H0}, the normal dispersion and pair amplitude are
\begin{align}
    \xi_{\bb k}&=-\mu+2t\bigl[\cos(\bb k\cdot\bb a_1)
       +\cos(\bb k\cdot\bb a_2)
       +\cos(\bb k\cdot(\bb a_2-\bb a_1))\bigr],
    \label{eq:trs_three_plus_dispersion}\\
    \Delta_{\bb k}&=-2\ii\Delta_{\rm p}\bigl[\sin(\bb k\cdot\bb a_1)
       +\sin(\bb k\cdot\bb a_2)
       +\sin(\bb k\cdot(\bb a_2-\bb a_1))\bigr],
    \qquad \xi_{\bb k}>|\Delta_{\bb k}|.
    \label{eq:trs_three_plus_pair_form_factor}
\end{align}
The thermal quadratic factor then damps large occupations, and the HS factors are unitary. Their finite-lattice products have convergent traces.

At zero temperature, the model poses a direct competition between pair correlations and single-boson condensation. In the absence of interactions, the odd-parity field pairs $b$ bosons near one valley with $c$ bosons near the other without producing a single-boson condensate. The triangular-lattice hopping has two minima at $\pm\bb K=\pm(4\pi/3,0)$. In the stable noninteracting ground state, $\langle\hat b_{\bb K}\rangle=\langle\hat c_{-\bb K}\rangle=0$, but $\langle\hat b_{\bb K}\hat c_{-\bb K}\rangle\ne0$. The atoms are correlated in pairs before they acquire individual phase coherence. A central zero-temperature question is whether interactions drive a transition to a single-boson condensate with both $\langle\hat b_{\bb K}\rangle\ne0$ and $\langle\hat c_{-\bb K}\rangle\ne0$, spontaneously breaking the flavor $SU(2)$ symmetry.

The same valley structure also determines the spatial pattern that emerges if both components condense. At a fixed total density $\rho$, shared equally by the two flavors, take the coherent-state amplitudes
\begin{equation}
    \langle\hat b_i\rangle=\sqrt{\rho/2}\,\ee^{\ii\bb K\cdot\bb r_i},
    \qquad
    \langle\hat c_i\rangle=-\ii\sqrt{\rho/2}\,\ee^{-\ii\bb K\cdot\bb r_i}.
    \label{eq:trs_odd_pair_counterflow}
\end{equation}
On a bond $j=i+\bb\delta$, the two flavor currents are $\rho\sin(\bb K\cdot\bb\delta)$ and $-\rho\sin(\bb K\cdot\bb\delta)$. Their sum vanishes, and the singlet amplitude $\langle\hat P_{ij}\rangle=-\rho\sin(\bb K\cdot\bb\delta)$ is real. In contrast, placing both flavors in the same valley gives a nonzero total current and zero singlet amplitude. Relative to that same-valley state, Eq.~\eqref{eq:trs_odd_pair_counterflow} lowers the coherent-state energy per site by $9g_1\rho^2+\sqrt3\Delta_{\rm p}\rho$. The current-square interaction and the pair field therefore both favor opposite flavor currents in this fixed-density comparison.

These opposite currents have a direct connection to frustrated magnetic order. In this coherent state, $\langle\hat b_i^+\hat c_i\rangle=-\ii(\rho/2)\ee^{-2\ii\bb K\cdot\bb r_i}$, whose phase rotates by $120^\circ$ between the three triangular sublattices. Interpreting $b,c$ as two spin states, this is a coplanar $120^\circ$ spin texture. Opposite-valley occupation and flavor counterflow also underlie the chiral spin superfluid studied in Ref.~\cite{liNatuParamekanti2014chiral}. Pairing further shifts the low-energy momenta through $E_{\bb k}=\sqrt{\xi_{\bb k}^2-|\Delta_{\bb k}|^2}$, opening the possibility of incommensurate spin textures. Spiral order with a continuously tunable wave vector has been realized in triangular optical-lattice simulators~\cite{struck2011frustrated}. This model links singlet pairing and strong interactions to the spatial organization of the condensate.

Beyond these zero-temperature properties, the model provides a setting to study the thermal melting of single-boson coherence and composite order. Theory predicts a chiral normal fluid between the superfluid and fully disordered phases in an orbital lattice model~\cite{liParamekantiHemmerich2014chiral}, whereas recent double-valley atomic experiments report the opposite sequence, with a time-reversal-symmetric superfluid as the intermediate phase~\cite{yuWuZhang2025vestigial}. These contrasting sequences expose a long-standing question: which order melts first, and when does an intermediate phase appear? This frustrated two-component model brings singlet pairing, single-boson coherence, and spin order into one sign-problem-free setting. Unbiased BAFQMC can determine their zero-temperature ordering and thermal evolution, resolve the competing melting scenarios, and identify intermediate phases.

\end{document}